# Conjugate heat transfer in the unbounded flow of a viscoelastic fluid past a sphere

F. Pimenta, M.A. Alves[*]

CEFT, Departamento de Engenharia Química, Faculdade de Engenharia da Universidade do Porto, Rua Dr. Roberto Frias, 4200-465 Porto, Portugal

**Abstract**

This work addresses the conjugate heat transfer of a simplified PTT fluid flowing past an unbounded sphere in the Stokes regime ($Re = 0.01$). The problem is numerically solved with the finite-volume method assuming axisymmetry, absence of natural convection and constant physical properties. The sphere generates heat at a constant and uniform rate, and the analysis is conducted for a range of Deborah ($0 \leq De \leq 100$), Prandtl ($10^0 \leq Pr \leq 10^5$) and Brinkman ($0 \leq Br \leq 100$) numbers, in the presence or absence of thermal contact resistance at the solid-fluid interface and for different conductivity ratios ($0.1 \leq \kappa \leq 10$). The drag coefficient shows a monotonic decrease with $De$, whereas the normalized stresses on the sphere surface and in the wake first increase and then decrease with $De$. A negative wake was observed for the two solvent viscosity ratios tested ($\beta = 0.1$ and $0.5$), being more intense for the more elastic fluid. In the absence of viscous dissipation, the average Nusselt number starts to decrease with $De$ after an initial increase. Heat transfer enhancement relative to an equivalent Newtonian fluid was observed for the whole range of conditions tested. The dimensionless temperature of the sphere decreases and becomes more homogeneous when its thermal conductivity increases in relation to the conductivity of the fluid, although small changes are observed in the Nusselt number. The thermal contact resistance at the interface increases the average temperature of the sphere, without affecting significantly the shape of the temperature profiles inside the sphere. When viscous dissipation is considered, significant changes are observed in the heat transfer process as $Br$ increases. Overall, a simplified PTT fluid can moderately enhance heat transfer compared to a Newtonian fluid, but increasing $De$ does not necessarily improve heat exchange.

[*] Corresponding author.
  Email address: mmalves@fe.up.pt (M.A. Alves)

# 1. Introduction

Heat transfer is a relevant process in several non-Newtonian fluid flows. From a theoretical standpoint, no flow is rigorously isothermal, although the isothermal assumption can be employed with negligible error in several situations of practical interest. Unless simplifications are undertaken, heat transfer processes often require the analysis of both solid and fluid phases, which interact between each other. In such conjugate heat transfer (CHT) problems, there is flow in the fluid and heat is exchanged between the different phases. The non-isothermal flow past a sphere is a CHT problem when the temperature distribution inside the sphere is non-trivial and results from the solution of the energy equation inside the solid domain. This occurs, for example, when a sphere generating heat in its interior is placed in a cold stream of fluid, which is the global picture of the problem addressed in the present work.

The non-isothermal flow past a bounded or unbounded sphere has been studied numerically and analytically for Newtonian (e.g. Kishore and Ramteke, 2016; Michaelides, 2006) and non-Newtonian fluids (Bhatnagar, 1970; Dhole et al., 2006a; Nirmalkar et al., 2014a; Nirmalkar et al., 2013a, b; Nirmalkar et al., 2014b; Prhashanna and Chhabra, 2010; Rajasekhar Reddy and Kishore, 2013; Sharma and Bhatnagar, 1975; Song et al., 2010, 2012; Westerberg and Finlayson, 1990). However, most of the studies with non-Newtonian fluids were limited to inelastic fluids, such as Power-law (Dhole et al., 2006a; Prhashanna and Chhabra, 2010; Rajasekhar Reddy and Kishore, 2013; Song et al., 2010, 2012), Bingham (Nirmalkar et al., 2014a; Nirmalkar et al., 2013b; Nirmalkar et al., 2014b) and Herschel-Bulkley (Nirmalkar et al., 2013a) fluids, and only a few addressed viscoelastic fluids (Bhatnagar, 1970; Sharma and Bhatnagar, 1975; Westerberg and Finlayson, 1990). These studies encompass natural, forced and mixed heat convection, for spheres which are either stationary or in rotation about a fixed axis. However, the focus was given to the fluid, whereas the sphere was simply modeled as a fixed temperature or fixed heat flux boundary condition. Among the few studies concerning viscoelastic fluids (Bhatnagar, 1970; Sharma and Bhatnagar, 1975; Westerberg and Finlayson, 1990), which date back to more than 30 years ago, the work by Westerberg and Finlayson (1990) explored a larger number of variables. They studied the heat transfer and flow past a sphere for different fluids (Newtonian, Generalized-Newtonian, Phan-Thien-Tanner and Upper-Convected Maxwell), assigning a fixed temperature to the sphere surface and keeping the Reynolds number below $10^{-3}$ (Westerberg and Finlayson, 1990). However, as recognized by Chhabra (2006), while the isothermal flow of viscoelastic fluids around spheres is relatively well known, there is still a generalized lack of knowledge on the non-isothermal flow of such fluids past a sphere, independently of the boundary conditions applied at the sphere surface.

In this work, we aim to reduce this gap of the literature by addressing numerically the conjugate heat transfer between an unbounded sphere and a simplified Phan-Thien-Tanner (PTT) fluid flowing over its surface. A constant and uniform volumetric heat source is prescribed at the interior of the sphere, which heats up the fluid in contact. Heat is transferred to the fluid by conduction and forced convection – natural convection is not considered. The system under study includes both the fluid and the sphere. The analysis is carried out for steady conditions, axisymmetric creeping flow ($Re$ = 0.01), varying Prandtl ($10^0 \leq Pr \leq 10^5$), Brinkman ($0 \leq Br \leq 100$) and Deborah ($0 \leq De \leq 100$) numbers, in the presence or absence of thermal contact resistance at the solid-fluid interface and for different ratios of the thermal conductivity between the solid and the fluid ($0.1 \leq \kappa \leq 10$). The two main objectives of this work are, firstly, to elucidate about the physics of the problem under study and, secondly, to provide benchmark data for a CHT problem involving a viscoelastic fluid, which, to the best of our knowledge, is among the first provided in the literature. Moreover, we also briefly discuss important aspects related with the application of a coupled-solution approach to CHT problems. Perhaps not less important is the release in open-source of the non-isothermal solver developed in this work (Pimenta and Alves, 2016), which can find applications in diverse and complex CHT problems involving non-Newtonian fluids (Agassant et al., 2017; Fernandes et al., 2016; Habla et al., 2016; Nóbrega et al., 2004a).

The remainder of this work is organized as follows. Section 2 describes the CHT problem under consideration in terms of geometry, mesh and boundary conditions. The governing equations are introduced in Section 3, and in Section 4 we present the finite-volume numerical method adopted to solve such equations. The dimensionless numbers controlling the CHT problem are identified in Section 5 and the main results obtained in this work are presented and discussed in Section 6. Finally, Section 7 ends the text with the main conclusions from this work.

**2. Problem description: geometry, mesh and boundary conditions**

Consider the unbounded flow of a viscoelastic fluid around a sphere which is generating heat at a constant and uniform rate in its interior. The geometry for this problem is schematically depicted in Fig. 1. The sphere with radius $R = D/2$ is immersed in a circular domain which extends up to 200$R$ around the sphere's center. The computational domain is made large enough such that the surrounding boundaries have a minimum impact on the solution (Appendix A). The outer boundaries of the fluid domain are kept circular to facilitate the meshing procedure and to obtain cells in the fluid domain with low non-orthogonality.

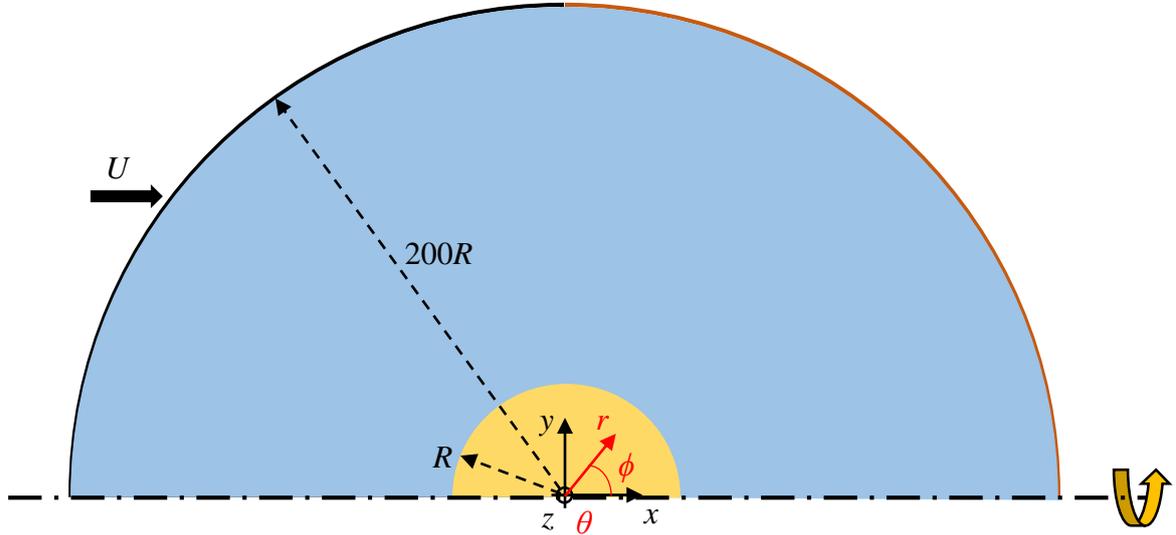

**Figure 1** – Geometry of the computational domain adopted for the numerical simulations (drawing not to scale). A sphere of radius $R$ is surrounded by a spherical fluid domain extending up to 200$R$ from the sphere's center. The dash-dotted line represents the axis of symmetry. Cartesian ($x,y,z$) and spherical ($r,\theta,\phi$) coordinate systems are represented.

Due to the flow axisymmetry around the *x*-axis, only a wedge of the total domain is effectively simulated, i.e. the meshes have a single cell in the $\theta$ direction. Such simplification does not compromise the accuracy in the range of dimensionless numbers simulated and drastically reduces the computational cost compared to a full 3D simulation.

Three meshes were used to assess the dependency of the numerical solution on spatial resolution (Appendix A). The main characteristics of the meshes are specified in Table 1, where $\Delta r_{min}$ and $\Delta \phi_{min}$ are the minimum cell spacing in the radial and azimuthal directions, respectively, and $n_{r,f}$ and $n_{\phi,f}$ are the number of cells in each of these directions, in the fluid domain. Cells are compressed toward the sphere surface in the radial direction and near the axis in the azimuthal direction. The meshes are composed by blocks in the solid and fluid domains, as shown in Fig. 2. Although optional, the distribution of cells was made perfectly continuous at the solid-fluid interface (Fig. 2).

**Table 1** – Characteristics of the computational grids used to assess the numerical accuracy with mesh refinement.

| Mesh | $\Delta r_{\min}/R = \Delta \phi_{\min}/R$ | $n_{\phi,\text{f}}$ | $n_{r,\text{f}}$ | Number of cells | |
|---|---|---|---|---|---|
| | | | | Fluid domain | Solid domain |
| M1 | 0.0032 | 300 | 234 | 70 200 | 32 200 |
| M2 | 0.0021 | 450 | 351 | 157 950 | 72 450 |
| M3 | 0.0016 | 600 | 468 | 280 800 | 128 800 |

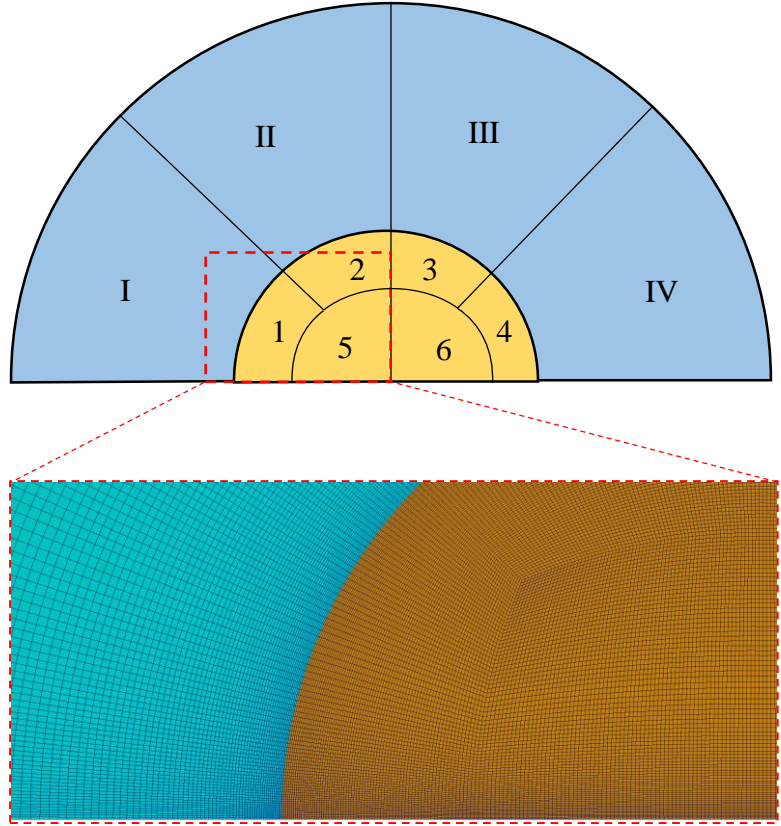

**Figure 2** – Distribution of blocks in the mesh (drawing not to scale). Roman numerals are used in the fluid domain and Arabic numerals are used in the solid domain. The distribution of blocks and cells is symmetric in relation to plane $x = 0$. The zoomed view at the bottom is for mesh M1.

The surface of the simulation domain was subdivided into five boundaries: the sphere surface, the axis of symmetry, the two sides of the wedge, a pseudo-inlet at $r = 200R \cap x \leq 0$ and a pseudo-outlet at $r = 200R \cap x > 0$. The following boundary conditions were applied:

- Sphere surface: $\mathbf{u} = \mathbf{0}$, $\nabla p \cdot \mathbf{n} = 0$, the components of $\boldsymbol{\tau}$ are linearly extrapolated (Pimenta and Alves, 2017) and the temperature on each side of the surface obeys Eqs. (6) and (7);
- Axis of symmetry: axial symmetry;

- Sides of the wedge: rotational periodicity;
- Inlet: $\mathbf{u} = (U, 0, 0)$, $\nabla p \cdot \mathbf{n} = 0$, $\boldsymbol{\tau} = \mathbf{0}$ and $T = T_0$;
- Outlet: $\nabla u_i \cdot \mathbf{n} = 0$, $p = 0$, $\nabla \tau_{ij} \cdot \mathbf{n} = 0$ and $\nabla T \cdot \mathbf{n} = 0$.

## 3. Governing equations

### 3.1. Fluid region

In the non-isothermal, incompressible, laminar flow of a viscoelastic fluid, the set of governing equations to be solved are mass conservation, momentum balance, energy conservation and the constitutive equation to evolve the polymeric extra-stresses. These equations are discussed next.

#### 3.1.1. Mass conservation and momentum balance

Mass conservation (Eq. 1) and momentum balance (Eq. 2) can be expressed as

$$\nabla \cdot \mathbf{u} = 0 \tag{1}$$

$$\rho_f \left( \frac{\partial \mathbf{u}}{\partial t} + \mathbf{u} \cdot \nabla \mathbf{u} \right) = -\nabla p + \eta_s \nabla^2 \mathbf{u} + \nabla \cdot \boldsymbol{\tau} \tag{2}$$

where $\mathbf{u}$ is the velocity vector, $t$ is the time, $p$ is the pressure, $\boldsymbol{\tau}$ is the polymeric extra-stresses tensor, $\rho_f$ is the fluid density and $\eta_s$ is the solvent viscosity. The total extra-stress tensor ($\boldsymbol{\tau}'$) is composed of a solvent contribution ($\boldsymbol{\tau}_s$) and a polymeric contribution ($\boldsymbol{\tau}$), such that $\boldsymbol{\tau}' = \boldsymbol{\tau}_s + \boldsymbol{\tau} = \eta_s (\nabla \mathbf{u} + \nabla \mathbf{u}^T) + \boldsymbol{\tau}$. The viscosity ($\eta_0$) is also split in solvent ($\eta_s$) and polymeric ($\eta_p$) contributions, such that $\eta_0 = \eta_s + \eta_p$ is the zero shear-rate viscosity. A Newtonian fluid is obtained for $\boldsymbol{\tau} = \mathbf{0}$ and $\eta_0 = \eta_s$.

#### 3.1.2. Constitutive equation

Several constitutive equations are available to model the rheology of viscoelastic fluids. In this study, we use the simplified PTT model (Phan-Thien and Tanner, 1977):

$$\exp\left( \frac{\varepsilon \lambda}{\eta_p} \text{tr}(\boldsymbol{\tau}) \right) \boldsymbol{\tau} + \lambda \stackrel{\nabla}{\boldsymbol{\tau}} = \eta_p (\nabla \mathbf{u} + \nabla \mathbf{u}^T) \tag{3}$$

where $\varepsilon$ is the extensibility parameter, $\lambda$ is the relaxation time of the fluid and $\stackrel{\nabla}{\boldsymbol{\tau}} = \frac{\partial \boldsymbol{\tau}}{\partial t} + \mathbf{u} \cdot \nabla \boldsymbol{\tau} - \boldsymbol{\tau} \cdot \nabla \mathbf{u} - \nabla \mathbf{u}^T \cdot \boldsymbol{\tau}$ represents the upper-convected time derivative of $\boldsymbol{\tau}$. A PTT fluid presents shear-thinning behavior in shear flow. The constant-viscosity Oldroyd-B model is recovered for $\varepsilon = 0$ and this model reduces to the Upper-Convected Maxwell model if additionally $\eta_s = 0$. Note that the PTT model, besides being realistic for several fluids (e.g. polymer melts), allows to reach significantly higher Deborah numbers than the Oldroyd-B model.

### 3.1.3. Energy equation

The conservation of energy, here expressed in terms of the temperature variable, can be written as (neglecting heat transfer by radiation)

$$\rho_f c_{p,f} \left( \frac{\partial T}{\partial t} + \mathbf{u} \cdot \nabla T \right) = \nabla \cdot \left( k_f \nabla T \right) + \boldsymbol{\tau}' : \nabla \mathbf{u} \tag{4}$$

where $T$ is the temperature, $k_f$ represents the thermal conductivity of the fluid and $c_{p,f}$ is the specific heat capacity of the fluid.

It is worth noting that the viscous dissipation term ($\boldsymbol{\tau}' : \nabla \mathbf{u}$) corresponds to the pure entropy elasticity case of the more general viscous dissipation term proposed by Peters and Baaijens (1997). Under such approximation, all mechanical energy is dissipated as heat (Wachs and Clermont, 2000). The results presented in this work for a Brinkman number of 0 were obtained by removing the viscous dissipation term from Eq. (4).

### 3.1.4. Temperature-dependent properties

In non-isothermal flows, the temperature affects the physical properties of the fluid. However, for small temperature variations the properties can be assumed approximately constant. In addition, some physical properties of some fluids are weakly sensitive to temperature changes. This is, for example, the case of the isobaric specific heat capacity of water, which presents a maximum variation of less than 1 % in the interval 0-100 °C. Moreover, considering temperature-dependent properties would introduce additional degrees of freedom to the thermal analysis and invalidate the one-way coupling discussed in Section 4.3. Therefore, we carried out the simulations assuming constant physical properties. In general, i.e. considering real viscoelastic fluids, while such approximation is acceptable for $\rho$, $c_p$ and $k$, some differences in the results shall be expected regarding $\eta$ and $\lambda$ (Bird et al., 1987; Nóbrega et al., 2004b). Note, however, that the numerical code used in this work is able to handle temperature-dependent physical properties (several laws are available) and the resulting non-linear coupling between equations.

## 3.2. Solid region

### 3.2.1. Energy equation

In the solid region, i.e. inside the sphere, only the energy equation is solved,

$$\rho_s c_{p,s} \frac{\partial T}{\partial t} = \nabla \cdot \left( k_s \nabla T \right) + \Phi \tag{5}$$

where $\rho_s$, $c_{p,s}$ and $k_s$ represent the density, specific heat capacity and thermal conductivity of the solid, respectively, and $\Phi$ is a volumetric energy source. The three physical properties are assumed temperature-independent and uniform in the whole sphere.

### 3.3. Solid-fluid interface

At the sphere surface, special care needs to be taken regarding the energy equation, which is solved on both sides of the interface (Eqs. 4 and 5). The conservation of energy imposes that the heat flux crossing the sphere surface is equal on both sides. Since at the sphere surface the heat is transferred by conduction, the previous condition is tantamount to

$$k_s \nabla T \big|_{s,i} \cdot \mathbf{n}_s = -k_f \nabla T \big|_{f,i} \cdot \mathbf{n}_f \Leftrightarrow -k_s \frac{\partial T}{\partial r}\bigg|_{r \to R^-} = -k_f \frac{\partial T}{\partial r}\bigg|_{r \to R^+} \quad (6)$$

If contact resistance is considered (Habla et al., 2016), a temperature jump is observed at the interface. In this general case, the temperature also obeys the condition

$$-k_s \nabla T \big|_{s,i} \cdot \mathbf{n}_s = h_{res}\left(T_{s,i} - T_{f,i}\right) \Leftrightarrow -k_s \frac{\partial T}{\partial r}\bigg|_{r \to R^-} = h_{res}\left(T_{s,i} - T_{f,i}\right) \quad (7)$$

where $T_{f,i}$ and $T_{s,i}$ represent the temperature at the interface on the fluid and solid sides, respectively, and $h_{res}$ is a heat transfer coefficient characterizing the contact resistance. The inverse of $h_{res}$ defines the thermal contact resistance, $R_C \equiv 1/h_{res}$. In order to keep the LHS of Eq. (7) defined, when $R_C \to 0$ then $\left(T_{s,i} - T_{f,i}\right) \to 0$, and this corresponds to perfect thermal contact.

### 4. Numerical method

The set of equations presented in the previous section was discretized and solved using rheoTool (Pimenta and Alves, 2016), an open-source toolbox implemented in the finite-volume framework of OpenFOAM®. The isothermal solvers available in the toolbox were modified in order to handle non-isothermal flows and multi-region domains in a coupled way. Only these two modifications will be discussed here. The details about the base viscoelastic fluid flow solver and coupled matrices were presented elsewhere (Pimenta and Alves, 2017, 2019). The resulting non-isothermal solver used to obtain the results presented in this work is freely available in rheoTool.

### 4.1. Solid-fluid temperature coupling

The numerical implementation of the boundary conditions in Eqs. (6) and (7) can assume several forms. When the energy equation for the solid and fluid phases are solved separately, a simple way is to derive explicit expressions for $T_{f,i}$ and $T_{s,i}$ from the discretized form of Eqs. (6) and (7). However, the resulting method is not accurate for transient simulations and may not converge in stiff cases. Imposing explicitly the flux, rather than the temperature itself, is also a possibility (Nóbrega et al., 2004a), but shares the same drawbacks as the previous approach. When a single energy equation including the multiple domains is solved, the method presented by Habla et al. (2016) can be applied. The energy equation is conditionally volume-averaged throughout the solid and fluid phases and the boundary conditions arise as an additional diffusive term at interface cells. This method has a simple implementation, but is only first-order accurate (Habla et al., 2016).

The approach implemented in this work is based on coupled matrices (Pimenta and Alves, 2019) and on the discretization of the diffusive term $\nabla \cdot (k \nabla T)$ for a space-varying conductivity (Patankar, 1980). Consider the 1D solid-fluid interface depicted in Fig. 3, where the boundary conditions of Eqs. (6) and (7) apply. It is assumed that the heat exchange between the two regions takes place by conduction and that a contact resistance ($1/h_{res}$) can exist at the interface – the temperature on each side of the interface may differ in that case ($T_{f,i} \neq T_{s,i}$). The question to be solved is how to express the conductivity at the interface, such that the diffusive term at the interface cells could be expressed as in any interior cell, i.e. as a function of ($T_s - T_f$)? According to Patankar (1980), equating the heat flux at the interface, as it appears in the discrete Laplace operator (using a Gaussian scheme), to the heat flux arising from a heat balance between points F and S (considering serial resistances) results in

$$k_i \frac{T_s - T_f}{\delta x_f + \delta x_s} = \frac{T_s - T_f}{\frac{\delta x_f}{k_f} + \frac{\delta x_s}{k_s} + \frac{1}{h_{res}}} \Rightarrow k_i = \frac{\delta x_f + \delta x_s}{\frac{\delta x_f}{k_f} + \frac{\delta x_s}{k_s} + \frac{1}{h_{res}}} \qquad (8)$$

In the absence of contact resistance, Eq. (8) is the well-known harmonic average of the conductivities on each side. The standard Laplace operator can be used at cells F and S, as long as the conductivity at the interface is computed with Eq. (8). It should be noted that Eq. (8) assumes that the area normal to the heat transfer direction is constant between points F and S, which is not true for the problem under study since the area increases in the radial direction. However, as the mesh is refined the distance between points F and S decreases and the area difference between these two points tends to zero. We confirmed numerically, in a one-dimensional case, that the convergence with mesh refinement is of second-order, even neglecting this small area variation. Therefore, Eq. (8) was used in the form it is presented.

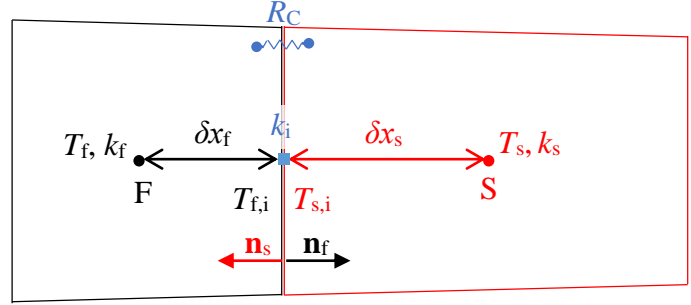

**Figure 3** – Schematic representation of two cells at the solid(s)-fluid(f) interface. The temperature and thermal conductivity are uniform in each cell, but the interface temperature ($T_{f,i}$ and $T_{s,i}$) can be different on each side of the interface due to the serial contact resistance ($R_C$).

In practice, the solution method starts with the assembling of the energy equation in each region. Each matrix of coefficients is then inserted in a coupled matrix (Pimenta and Alves, 2019) and the coefficients of the Laplace operator at interface cells are then modified according to Eq. (8). These coefficients ensure the coupling between the two energy equations. If both the solid and fluid regions are composed exclusively of orthogonal cells and the time-derivatives are removed from the equations, the solution of the coupled matrix is obtained in a single iteration. This contrasts with the multiple iterations needed by the methods referred above to obtain the steady-state temperature field in both regions (Habla et al., 2016; Nóbrega et al., 2004a). Moreover, since the coupling between regions is only at the matrix level, there is no need for the duplication of fields and/or meshes used in some methods (Weber et al., 2019).

Although the temperature on each side of the interface ($T_{f,i}$ and $T_{s,i}$) could be determined based on the heat flux crossing the interface (equal on both sides), we instead derived expressions for each temperature based on the discretized form of Eqs. (6) and (7),

$$\begin{cases} k_s \dfrac{T_{s,i} - T_s}{\delta x_s} = -k_f \dfrac{T_{f,i} - T_f}{\delta x_f} \\ -k_s \dfrac{T_{s,i} - T_s}{\delta x_s} = h_{res}(T_{s,i} - T_{f,i}) \end{cases} \quad (9)$$

After some basic algebraic manipulation of these expressions, which form a system of two equations on two unknowns ($T_{f,i}$ and $T_{s,i}$), we arrive at

$$T_{s,i} = \dfrac{h_{res}(T_s k_s \delta x_f + T_f k_f \delta x_s) + T_s k_s k_f}{h_{res}(k_s \delta x_f + k_f \delta x_s) + k_s k_f} \quad (10)$$

$$T_{f,i} = \frac{T_{s,i} h_{res} \delta x_f + T_f k_f}{h_{res} \delta x_f + k_f} \tag{11}$$

In the limit of no contact resistance ($R_C \to 0$), i.e. perfect thermal contact, the temperature is continuous across the interface,

$$T_{s,i} = T_{f,i} = T_i = \frac{T_s k_s \delta x_f + T_f k_f \delta x_s}{k_s \delta x_f + k_f \delta x_s} \tag{12}$$

It should be noted that Eqs. (9), (10) and (11) could be used instead of the interpolated thermal conductivity to impose the implicit coupling between the energy equation in the two regions.

In Fig. 3, the two cells represented have a matching face. However, the solver was implemented in such a way that non-matching faces are also allowed, using the so-called arbitrary mesh interface (AMI) available in OpenFOAM®. In practice, area-weighted matrix coefficients are introduced to allow such interfaces.

### 4.2. The log-conformation tensor approach for non-isothermal flows

The log-conformation tensor approach was proposed by Fattal and Kupferman (2004) to mitigate the high Weissenberg number problem in isothermal flows, i.e. for constant viscosity coefficients and relaxation time. This technique was used in the present work without modifications, since we do not consider temperature-dependent properties. However, it is important to note, as general case, that when the viscosity and relaxation time are temperature-dependent and their ratio $\frac{\eta_p(T)}{\lambda(T)}$ is not constant, the log-conformation tensor approach may need to be modified. It can be shown that an additional term involving the material derivative $\frac{D}{Dt}\left(\frac{\lambda(T)}{\eta_p(T)}\right)$ needs to be taken into account to ensure the analytical equivalence between the $\tau$-based and the conformation-tensor-based constitutive equation.

### 4.3. Discretization and solution method

When the physical properties of the fluid are independent of temperature, as considered in this work, there is a one-way coupling between flow and heat transfer. The flow affects the heat transfer through the viscous dissipation and convective terms in the energy equation (Eq. 4), but the temperature field has no effect on the flow. In such cases, it is possible to decouple hydrodynamics from heat transfer, which, in practice, allows simulating different heat transfer scenarios from a single set of *p*-**u**-**τ** fields (Kishore and Ramteke, 2016). Therefore, the simulations are split into two stages. Firstly, we seek to solve the flow for a given set of (*Re*, *De*, *β*, *ε*) values, disregarding heat transfer. Secondly, we evaluate several heat transfer scenarios for the computed *p*-**u**-**τ** fields.

The SIMPLEC algorithm (van Doormaal and Raithby, 1984) for segregated solution methods is applied for pressure-velocity coupling when solving for the flow, along with the technique presented by Pimenta and Alves (2017) to couple polymeric stresses and velocity. The method is inherently transient (Pimenta and Alves, 2017), but steady-state solutions are reached upon time evolution for steady flows. Since only steady-state solutions are sought in this work, the order of accuracy of the schemes used to discretize time-derivatives is of minor importance. A first-order implicit Euler scheme was employed for this purpose. The dimensionless time-step used in the flow simulations was $\Delta \tilde{t} = 0.0025$ and the simulations converged around $\tilde{t} \approx 200$. As noted by Harlen (2002), the steadiness of the drag coefficient should not be used alone to assess convergence, since it typically stabilizes before the stresses in the wake of the sphere. Convective terms were discretized with the CUBISTA high-resolution scheme (Alves et al., 2003) and the Green-Gauss theorem was used to compute gradients and derivatives (except $\nabla \mathbf{u}$, computed with a least-squares method), where linear interpolation was applied to interpolate values from cell centers to face centers.

The coupled solution method to solve the energy equation was described in detail in Pimenta and Alves (2019). The enhanced stability of the coupled solver allows the removal of time-derivatives from the equation. Convergence was typically achieved in less than 20 iterations depending on the conditions (more than one iteration is needed to converge the simulations due to the mesh non-orthogonality in the solid domain). The schemes mentioned for the flow equations were also used to discretize the convective and diffusive terms of the energy equation in the solid and fluid regions.

## 5. Dimensionless numbers

The governing equations can be rendered dimensionless using the following set of normalized variables (written with a tilde): $\tilde{t} = \dfrac{tU}{D}$ , $\tilde{x} = \dfrac{x}{D}$ , $\tilde{\mathbf{u}} = \dfrac{\mathbf{u}}{U}$ , $\tilde{\boldsymbol{\tau}} = \dfrac{\boldsymbol{\tau} D}{\eta_0 U}$ , $\tilde{p} = \dfrac{pD}{\eta_0 U}$ and $\tilde{T} = \dfrac{T - T_0}{\bar{q}_w D / k_f} = 6 k_f \dfrac{T - T_0}{\Phi D^2}$ . In the case of the dimensionless temperature, the surface-averaged (non-uniform) heat flux across the interface $\left(\bar{q}_w = \dfrac{\Phi D}{6}\right)$ is used in the normalization.

Once the dimensionless variables are replaced in the governing equations, a set of dimensionless numbers arise and completely define the problem under study. The dimensionless numbers governing the flow in our study are the Reynolds number, $Re = \dfrac{\rho_f UD}{\eta_0}$, the Deborah number, $De = \dfrac{\lambda U}{D}$, the solvent viscosity ratio, $\beta = \dfrac{\eta_s}{\eta_0}$, and the extensibility parameter of the PTT model ($\varepsilon$, see Eq. 3). On the other hand, the thermal component of the problem is controlled by the Prandtl number, $Pr = \dfrac{\eta_0 c_{p,f}}{k_f}$, the Péclet number, $Pe = Re\,Pr = \dfrac{\rho_f UD c_{p,f}}{k_f}$, which arises as the product of two other dimensionless numbers, the Brinkman number, $Br = \dfrac{6\eta_0}{\Phi}\left(\dfrac{U}{D}\right)^2$, the thermal conductivity ratio, $\kappa = \dfrac{k_s}{k_f}$, and the dimensionless contact resistance, $\Omega = \dfrac{k_f}{h_{res} D}$.

The Nusselt number, Biot number, volume-averaged temperature of the sphere and drag coefficient are dimensionless quantities of interest in the post-processing stage. The Nusselt number corresponds to the dimensionless form of the heat transfer coefficient defined at the sphere surface (fluid side). It varies in the azimuthal direction, such that a local Nusselt number can be defined as

$$Nu_\phi = \dfrac{D}{(T_{w,\phi} - T_0)}\left(-\dfrac{\partial T}{\partial r}\right)_{w,\phi} \tag{13}$$

where subscript $_{w,\phi}$ denotes a variable evaluated at a given azimuth on the sphere surface, in the fluid side. A surface-averaged Nusselt number ($\overline{Nu}$) can be also defined by integrating $Nu_\phi$ over the sphere surface ($S_s$),

$$\overline{Nu} = \dfrac{1}{S_s}\int_{S_s} Nu_\phi \, dS \tag{14}$$

Note that $T_0$ in Eq. (13) should be replaced by a bulk temperature to keep the definition consistent with an energy balance at the sphere surface. The definition adopted for $Nu_\phi$, though not rigorous from that standpoint, is widely used in the literature and easy to apply. The occurrence of locally negative $Nu_\phi$ values in this work is a consequence of this choice.

The Biot number ($Bi$) applies to the problem under analysis due to the occurrence of temperature gradients inside the sphere, and represents the ratio between the thermal resistance inside the sphere and the external thermal resistance. The expression for the Biot number is given by

$$Bi = \frac{\dfrac{L}{k_s}}{\dfrac{1}{h}+\dfrac{1}{h_{res}}} = \frac{hh_{res}D}{6k_s(h+h_{res})} = \frac{h_{res}}{(h+h_{res})}\frac{\overline{Nu}}{6\kappa} \tag{15}$$

where $L$ is a characteristic length representing the ratio between the volume and surface of the sphere, i.e. $L = D/6$ and $h = k_f \overline{Nu}/D$ is the convective heat transfer coefficient. Note that at the sphere surface the contact resistance ($1/h_{res}$) is in series with $1/h$. In the absence of contact resistance $\left(\dfrac{1}{h_{res}} \to 0\right)$, we simply have $Bi = \dfrac{\overline{Nu}}{6\kappa}$.

The volume-averaged temperature of the sphere corresponds to the integral of the temperature field over the sphere volume, normalized by the total volume ($V_s$),

$$\overline{\tilde{T}_s} = \frac{1}{V_s}\int_{V_s} \tilde{T} dV \tag{16}$$

The drag coefficient ($C_d$) represents the component of the force exerted by the fluid on the sphere, in the flow direction $\hat{\mathbf{i}} = (1,0,0)$, normalized by the dynamic pressure $\left(\dfrac{\rho_f U^2}{2}\right)$ and projected area, and is given by

$$C_d = \frac{8}{\rho_f U^2 \pi D^2}\int_{S_s}\left[-p\mathbf{I}+\eta_s\left(\nabla\mathbf{u}+\nabla\mathbf{u}^T\right)+\boldsymbol{\tau}\right]\cdot\hat{\mathbf{n}}\cdot\hat{\mathbf{i}}\,dS \tag{17}$$

Note, however, that the results for the drag coefficient in the next section will be presented as the product $C_d Re$.

## 6. Results and discussion

Before applying the non-isothermal solver to the problem under study, several verification tests were carried out to ensure the correctness of the algorithm. Among other tests, we verified the analytical solution for the heat transfer between two slabs of different conductivities with/without contact resistance at the interface (Nóbrega et al., 2004a), the numerical solution for the conjugate heat transfer between a calibrator and a polymer layer (Nóbrega et al., 2004a) and the numerical solution for heat transfer in the laminar flow of a simplified PTT fluid in a pipe (Nóbrega et al., 2004b). The solver showed a good agreement with the available analytical/numerical solutions in all the tested cases. The results are not shown for conciseness, but some of these tests have been made available as tutorials in rheoTool.

All the simulations were performed in the laminar (Stokesian) flow regime, for $Re = 0.01$, which is representative of viscous fluids and/or small-scale flows. The low $Re$ limit has received relatively less attention than the mid-high $Re$ range. The extensibility parameter of the PTT model was also kept fixed at $\varepsilon = 0.25$, which in practice represents concentrated polymers or polymer melts; this is a common value used in the literature (Afonso et al., 2008; Jin et al., 1991; Sun and Tanner, 1994; Zheng et al., 1991). The remaining parameters ($Pr$, $\kappa$, $\Omega$ and $Br$) were varied independently and an individual section is dedicated to the effect of each one for different values of $De$ and two values of $\beta$ (0.1 and 0.5). These two values of $\beta$ are high for polymer melts, but can reproduce viscoelastic fluids typically used in thermal studies at small scales (Abed et al., 2016; Whalley et al., 2015).

Before proceeding to the thermal analysis, we first characterize the flow around the sphere in the next section.

**6.1. Flow characteristics**

The drag coefficient for the range of $De$ tested is presented in Table 2 and plotted in Fig. 4. The drag coefficient decreases monotonically with $De$ due to shear-thinning and seems to approach an horizontal asymptote at high $De$. The horizontal dashed lines plotted in Fig. 4 correspond to the drag coefficient for the Newtonian component of the fluid, $C_d Re = 24\beta\left(1 + \frac{3}{16}Re\right)$, assuming Oseen's drag coefficient, $C_d Re = 24\left(1 + \frac{3}{16}Re\right)$, for Newtonian fluids at $Re \leq 1$ (Dey et al., 2019). The results suggest that the polymeric contribution to the drag coefficient decreases as $De$ increases. A natural consequence of this hypothesis is that the drag coefficient is lower for more elastic fluids, i.e. lower $\beta$ values, which is confirmed by the data in Table 2 and Fig. 4. This is in contrast with the results obtained from a few tests carried out with an Oldroyd-B fluid ($\beta = 0.5$; Fig. 5 and Table 2). For this constant-viscosity viscoelastic fluid, the drag coefficient shows an initial decrease up to $De \approx 1$, followed by a small increase with $De$, apparently showing a positive contribution of elasticity to $C_d$. These results for the Oldroyd-B fluid are in reasonable agreement with the correlation recently proposed by Faroughi et al. (2020) (plotted in Fig. 5), showing a maximum error of ~ 1 % for the set of *De* tested.

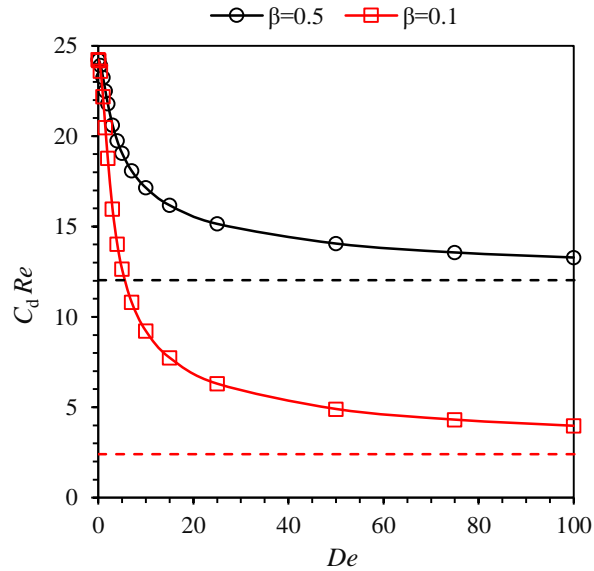

**Figure 4** – Variation of the drag coefficient as a function of *De*. The dashed lines correspond to $C_d Re = 24\beta\left(1+\frac{3}{16}Re\right)$. The solid lines are only a guide to the eye.

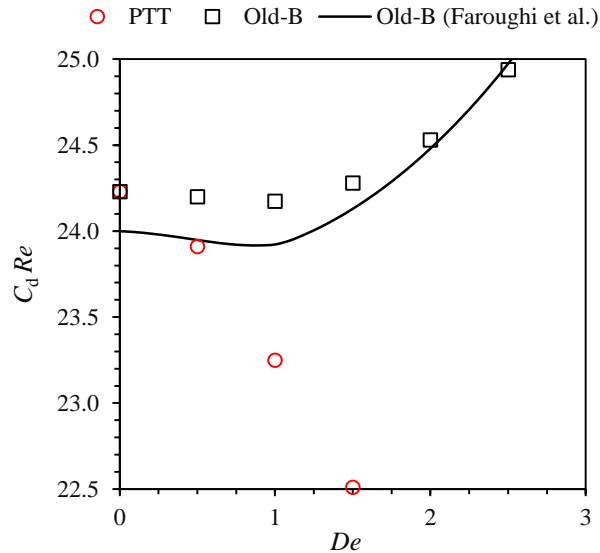

**Figure 5** – Comparison of the drag coefficient obtained for a PTT fluid and an Oldroyd-B fluid ($\beta = 0.5$ in both cases). The solid line represents the correlation by Faroughi et al. (2020) for an Oldroyd-B fluid in similar conditions.

The velocity and the total stress components aligned with the axis of symmetry are plotted in Fig. 6 for the PTT fluid. The profiles are taken along the axis of the simulation domain ($x/R < -1$ and $x/R > 1$) and over the sphere surface ($-1 \leq x/R \leq 1$). The velocity profiles for the lowest $\beta$ value show the existence of a negative wake, i.e. a region of fluid in the wake of the sphere moving in the opposite direction to the sphere (considering the sphere moving in a fluid at rest) (Arigo and McKinley, 1998; Harlen, 2002). In the laboratory frame of reference, this is seen as the region of the velocity profile for which $\tilde{u}_x > 1$. The negative wake is also present for $\beta = 0.5$, but at a lower extent, as shown in the insets of Figs. 6(a) and (b), containing the maximum axial velocity values reached at each $De$. Regarding the negative wake, the fluid with lower $\beta$ starts forming the wake at a lower $De$ and the maximum velocity in the wake is higher, but also starts decaying at lower $De$ ($De \approx 25$ for $\beta = 0.1$ and $De \approx 50$ for $\beta = 0.5$). Moreover, the negative wake seems to start farther apart from the sphere for increasing $De$, while also extending over a longer region. For the $De$ values not showing a negative wake, it is still interesting to note that the velocity in the wake recovers faster than in the Newtonian case. This is opposite to what is observed for constant-viscosity viscoelastic fluids, which present a slower decay (extended wake) as a result of strain-hardening in the wake of the sphere (Harlen, 2002).

**Table 2** – Drag coefficient for the unbounded flow of PTT and Oldroyd-B fluids around a sphere.

| $De$ | $C_dRe$ (PTT) | | $De$ | $C_dRe$ (Oldroyd-B) |
|---|---|---|---|---|
| | $\beta = 0.1$ | $\beta = 0.5$ | | $\beta = 0.5$ |
| 0 | 24.230 | 24.230 | 0.5 | 24.200 |
| 0.1 | 24.190 | 24.210 | 1 | 24.174 |
| 0.5 | 23.600 | 23.910 | 1.5 | 24.280 |
| 1 | 22.180 | 23.250 | 2 | 24.530 |
| 1.5 | 20.470 | 22.510 | 2.5 | 24.940 |
| 2 | 18.770 | 21.800 | | |
| 3 | 15.970 | 20.620 | | |
| 4 | 14.020 | 19.740 | | |
| 5 | 12.640 | 19.060 | | |
| 7 | 10.820 | 18.090 | | |
| 10 | 9.220 | 17.146 | | |
| 15 | 7.740 | 16.180 | | |
| 25 | 6.300 | 15.150 | | |
| 50 | 4.900 | 14.060 | | |
| 75 | 4.310 | 13.570 | | |
| 100 | 3.970 | 13.290 | | |

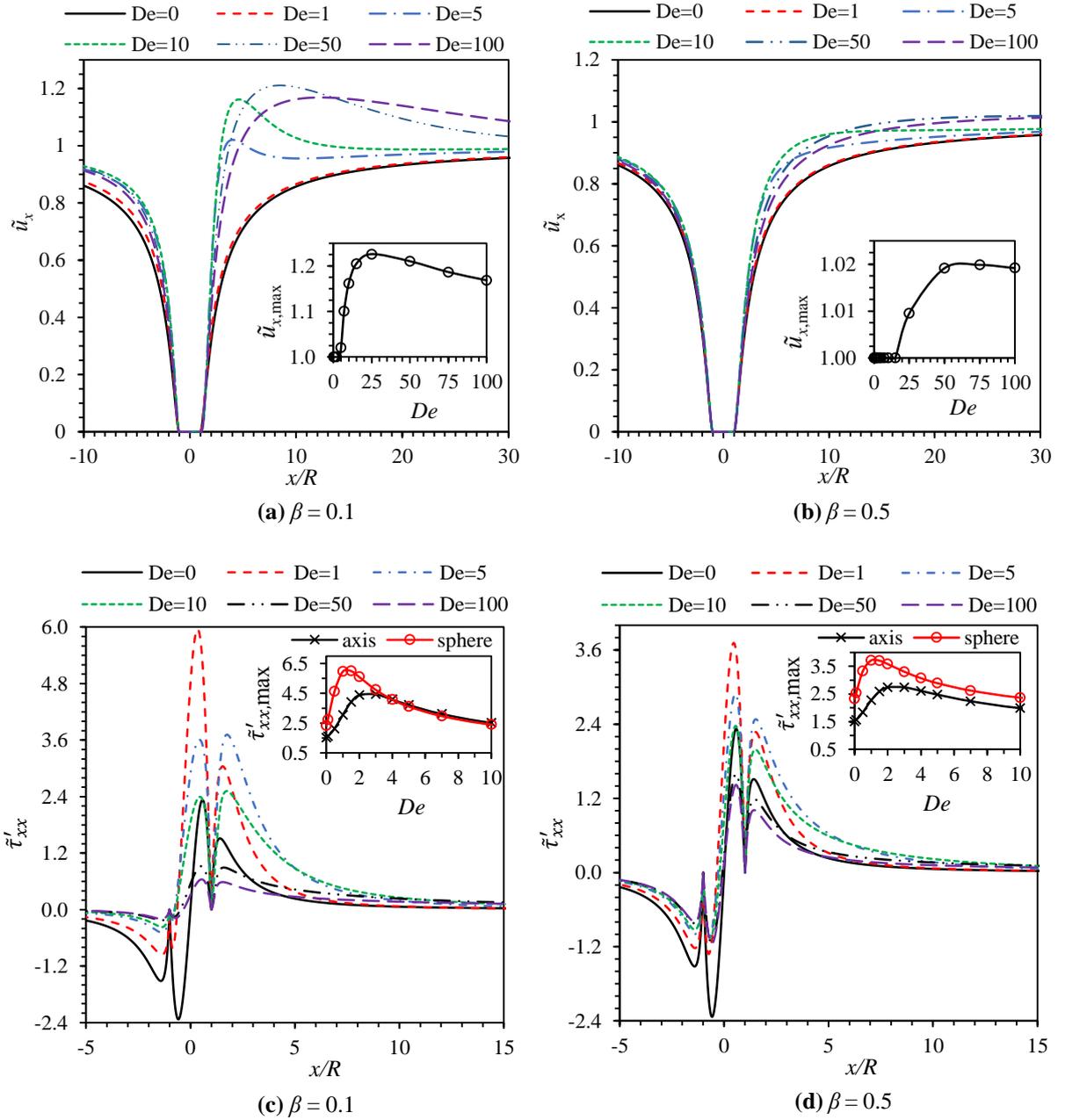

**Figure 6** – Velocity (a and b) and total stresses (c and d) profiles. The profiles are taken at the axis for $x/R < -1$ and $x/R > 1$, whereas they are sampled from the sphere surface for $-1 \leq x/R \leq 1$. The insets in (a) and (b) represent the maximum axial velocity as a function of $De$ (the line is a guide to the eye). The insets in (c) and (d) contain the maximum $\tilde{\tau}'_{xx}$ component value over the sphere surface (circle symbols) and axis (cross symbols) for different $De$ (the lines are a guide to the eye).

The stress profiles in Figs. 6(c) and (d) display a minimum upstream of the sphere, before they vanish at the stagnation point $x/R = -1$. At the sphere surface, the profiles are bell-shaped and a peak value is reached for $\phi < 90º$. A second peak is observed in the wake of the sphere, less than one radius downstream of the sphere. The insets in Figs. 6(c) and (d) show the maximum $\tilde{\tau}'_{xx}$ for each $De$, at the sphere surface and in the wake. Firstly, it can be seen that $\tilde{\tau}'_{xx,\max}$ increases up to a critical *De* and decays beyond that point. The critical *De* is, however, different for the sphere surface and the wake. Secondly, after the critical *De* is reached, the difference between the two curves decreases, whereas it increases before the critical *De*, with the peak at the sphere surface being higher than the peak in the wake.

Notwithstanding the non-monotonic behavior of the stress profiles with *De*, the drag coefficient shows a monotonic decrease with *De*, mostly due to shear-thinning and the dominant shear contribution to the drag coefficient.

The contours of velocity and extra-stresses are plotted in Fig. 7 for *De* = 2 and *De* = 25, and the two values of $\beta$. The negative wake can be clearly seen in the velocity contours for *De* = 25 and $\beta$ = 0.1. These contours further show that all components of the polymeric stress tensor are higher (in absolute value) for *De* = 2 than for *De* = 25, being also higher for the more elastic fluid ($\beta$ = 0.1). It is therefore expectable that the behavior seen in the insets of Figs. 6(a) and (b) for $\tilde{\tau}'_{xx}$ can be extended to the remaining components of the polymeric stress tensor.

The results discussed above are consistent with related works in the literature concerning PTT fluids. In fact, Afonso et al. (2008) studied the flow past a confined cylinder and reported a monotonic decrease of the drag coefficient with *De*, the existence of a negative wake (more intense for lower values of $\varepsilon$) and a similar trend for the stresses on the cylinder surface (an initial increase with *De*, followed by a decrease above a critical *De*; for $\varepsilon$ = 0.25 and the linearized form of a PTT fluid). The monotonic decrease of the drag coefficient with *De* and the presence of a negative wake was also verified by Jin et al. (1991), Sun and Tanner (1994) and Zheng et al. (1991) for the confined flow past a sphere.

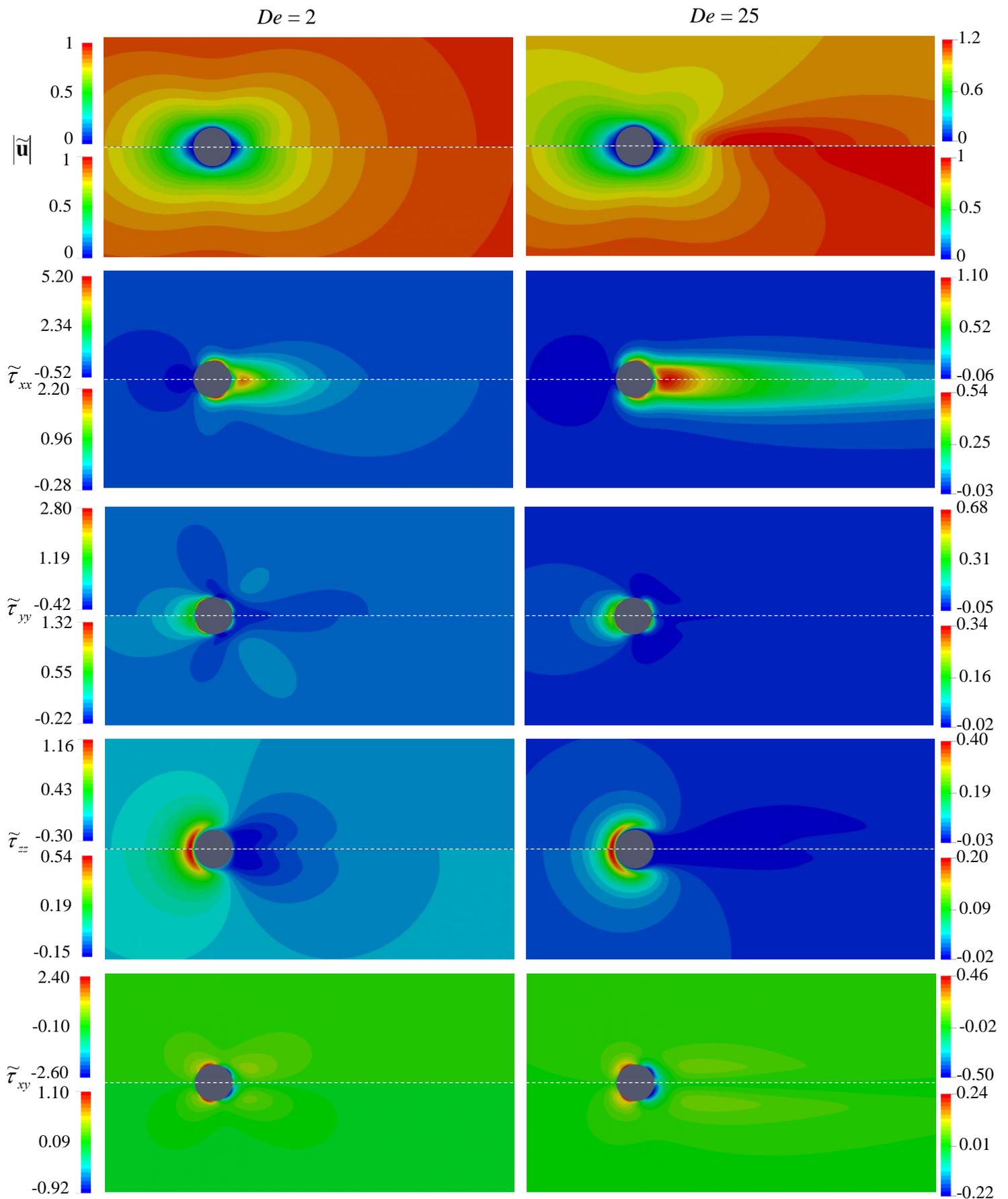

**Figure 7** – Contours of velocity and polymeric stresses for $De = 2$ (left) and $De = 25$ (right). In each subfigure, the upper part is for $\beta = 0.1$ and the lower part is for $\beta = 0.5$. The flow is from left to right.

## 6.2. Heat transfer

The temperature profiles over the sphere surface presented in the following sections are equal on the solid and fluid side when $\Omega = 0$. In such cases, we will not differentiate between both sides of the sphere.

### 6.2.1. Prandtl number effect

The results presented in this section are for fixed $Br = 0$, $\kappa = 1$ and $\Omega = 0$. With the purpose of examining the effect of $Pr$ for different $De$ and $\beta$, the CHT problem was solved for $0 \leq De \leq 100$, $1 \leq Pr \leq 10^5$ (corresponding to $10^{-2} \leq Pe \leq 10^3$) and $\beta = 0.1, 0.5$. The range of $De$ covers from Newtonian to highly elastic flows and the range of $Pr$ (and $Pe$) extends from conduction-dominated to convection-dominated heat transfer. The numerical stability was not an issue, even near the upper boundary of those ranges, but accuracy starts to be an issue of concern beyond those limits.

The surface-averaged Nusselt number is presented in Table 3 and the volume-averaged temperature of the sphere is shown in Table 4. The surface-averaged Nusselt number is also plotted in Fig. 8 for $\beta = 0.1$ as the ratio between the value at a given $De$ and the value obtained for a Newtonian fluid. This ratio is above 1 for all the conditions tested and the fluid with lower $\beta$ (more elastic) presents higher values of $\overline{Nu}$. Thus, viscoelasticity enhances heat transfer, but in a non-monotonic manner. Indeed, similarly to the stress profiles, the Nusselt number increases up to a critical $De$ and then decreases. In the range of conditions tested, this critical $De$ is always higher for the lower $\beta$ tested and tends to decrease as $Pr$ increases. For example, the critical $De$ is approximately 10–15 at $Pr = 10^2$ and 5–7 at $Pr = 10^5$. These values of critical $De$ are higher than those observed for the stress profiles (inset of Figs. 6c and d), but smaller than the $De$ at which the maximum wake velocity is reached (inset of Figs. 6a and b). Therefore, a direct cause-effect relationship can not be established, although it is clear that the inversion of behavior in the $\overline{Nu} - De$ relation is related to the rearrangement of the velocity field around the sphere, via the convective term in the energy equation, which is itself related to the polymeric stresses. The maximum $De$ reached in the simulations is not high enough to disclose the asymptotic Nusselt number to which the profiles seem to tend for increasing $De$. The increase of the Nusselt number observed at low $De$ is consistent with the results of Westerberg and Finlayson (1990), where the same behavior was reported up to $De = 0.7$ (the maximum $De$ reached in that work for a PTT fluid with $\varepsilon = 0.015$, $\eta_s/\eta_p = 1/8$ and $Pe \leq 100$).

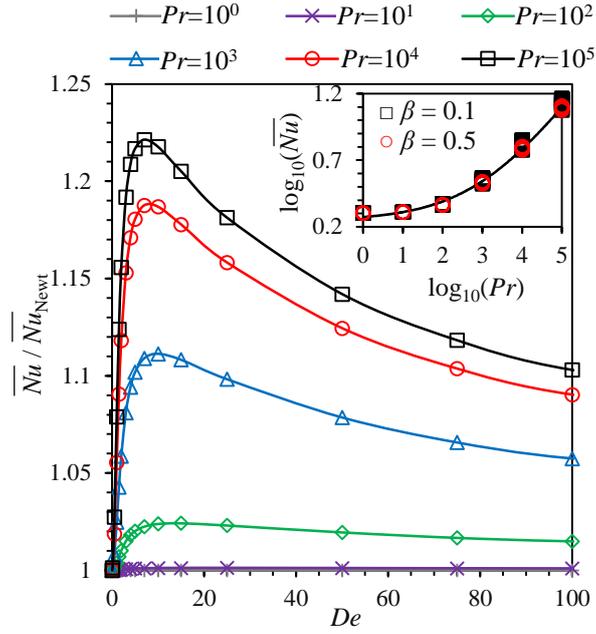

**Figure 8** − Variation of the surface-averaged Nusselt number as a function of $De$ for several $Pr$ ($\beta = 0.1$, $Br = 0$, $\kappa = 1$ and $\Omega = 0$). The Nusselt number is plotted as the ratio between $\overline{Nu}$ at a given $De$ and $\overline{Nu}$ for $De = 0$, representing the heat transfer enhancement relative to the Newtonian case. The lines are a guide to the eye. The inset plots $\log_{10}(\overline{Nu})$ as a function of $\log_{10}(Pr)$ for $\beta = 0.1$ and $\beta = 0.5$, and all the $De$ tested. The line represents a power-law fit to the data (see the text).

**Table 3** – Surface-averaged Nusselt number ($\overline{Nu}$) for different $De$, $\beta$ and $Pr$ ($Br = 0$, $\kappa = 1$ and $\Omega = 0$).

| De | $Pr = 10^0$ | | $Pr = 10^1$ | | $Pr = 10^2$ | | $Pr = 10^3$ | | $Pr = 10^4$ | | $Pr = 10^5$ | |
|---|---|---|---|---|---|---|---|---|---|---|---|---|
| | $\beta = 0.1$ | $\beta = 0.5$ | $\beta = 0.1$ | $\beta = 0.5$ | $\beta = 0.1$ | $\beta = 0.5$ | $\beta = 0.1$ | $\beta = 0.5$ | $\beta = 0.1$ | $\beta = 0.5$ | $\beta = 0.1$ | $\beta = 0.5$ |
| 0 | 2.007 | 2.007 | 2.044 | 2.044 | 2.307 | 2.307 | 3.336 | 3.336 | 6.002 | 6.002 | 11.956 | 11.956 |
| 0.1 | 2.007 | 2.007 | 2.044 | 2.044 | 2.307 | 2.307 | 3.337 | 3.337 | 6.008 | 6.006 | 11.971 | 11.965 |
| 0.5 | 2.007 | 2.007 | 2.044 | 2.044 | 2.309 | 2.308 | 3.362 | 3.349 | 6.114 | 6.059 | 12.284 | 12.118 |
| 1 | 2.007 | 2.007 | 2.045 | 2.045 | 2.316 | 2.311 | 3.419 | 3.373 | 6.336 | 6.147 | 12.900 | 12.348 |
| 1.5 | 2.007 | 2.007 | 2.045 | 2.045 | 2.323 | 2.314 | 3.479 | 3.395 | 6.546 | 6.215 | 13.437 | 12.508 |
| 2 | 2.007 | 2.007 | 2.045 | 2.045 | 2.330 | 2.317 | 3.532 | 3.412 | 6.712 | 6.262 | 13.818 | 12.608 |
| 3 | 2.007 | 2.007 | 2.046 | 2.045 | 2.341 | 2.320 | 3.606 | 3.433 | 6.921 | 6.312 | 14.249 | 12.706 |
| 4 | 2.007 | 2.007 | 2.046 | 2.045 | 2.349 | 2.322 | 3.650 | 3.443 | 7.029 | 6.333 | 14.451 | 12.740 |
| 5 | 2.007 | 2.007 | 2.046 | 2.045 | 2.353 | 2.324 | 3.676 | 3.448 | 7.086 | 6.340 | 14.548 | 12.748 |
| 7 | 2.007 | 2.007 | 2.047 | 2.045 | 2.359 | 2.325 | 3.700 | 3.450 | 7.129 | 6.338 | 14.603 | 12.732 |
| 10 | 2.007 | 2.007 | 2.047 | 2.045 | 2.362 | 2.326 | 3.707 | 3.447 | 7.124 | 6.321 | 14.559 | 12.685 |
| 15 | 2.007 | 2.007 | 2.047 | 2.045 | 2.363 | 2.325 | 3.697 | 3.438 | 7.070 | 6.289 | 14.409 | 12.607 |
| 25 | 2.007 | 2.007 | 2.047 | 2.045 | 2.360 | 2.323 | 3.664 | 3.421 | 6.952 | 6.239 | 14.123 | 12.490 |
| 50 | 2.007 | 2.007 | 2.047 | 2.045 | 2.352 | 2.319 | 3.598 | 3.397 | 6.749 | 6.168 | 13.654 | 12.330 |
| 75 | 2.007 | 2.007 | 2.046 | 2.045 | 2.345 | 2.316 | 3.555 | 3.383 | 6.625 | 6.132 | 13.370 | 12.248 |
| 100 | 2.007 | 2.007 | 2.046 | 2.045 | 2.341 | 2.315 | 3.528 | 3.375 | 6.545 | 6.109 | 13.188 | 12.197 |

**Table 4** – Volume-averaged sphere temperature ($\overline{\overline{T}}_s$) for different $De$, $\beta$ and $Pr$ ($Br = 0$, $\kappa = 1$ and $\Omega = 0$).

| De | $Pr = 10^0$ | | $Pr = 10^1$ | | $Pr = 10^2$ | | $Pr = 10^3$ | | $Pr = 10^4$ | | $Pr = 10^5$ | |
|---|---|---|---|---|---|---|---|---|---|---|---|---|
| | $\beta = 0.1$ | $\beta = 0.5$ | $\beta = 0.1$ | $\beta = 0.5$ | $\beta = 0.1$ | $\beta = 0.5$ | $\beta = 0.1$ | $\beta = 0.5$ | $\beta = 0.1$ | $\beta = 0.5$ | $\beta = 0.1$ | $\beta = 0.5$ |
| 0 | 0.598 | 0.598 | 0.589 | 0.589 | 0.534 | 0.534 | 0.403 | 0.403 | 0.272 | 0.272 | 0.187 | 0.187 |
| 0.1 | 0.598 | 0.598 | 0.589 | 0.589 | 0.534 | 0.534 | 0.403 | 0.403 | 0.272 | 0.272 | 0.187 | 0.187 |
| 0.5 | 0.598 | 0.598 | 0.589 | 0.589 | 0.533 | 0.534 | 0.401 | 0.402 | 0.269 | 0.270 | 0.185 | 0.186 |
| 1 | 0.598 | 0.598 | 0.589 | 0.589 | 0.532 | 0.533 | 0.396 | 0.400 | 0.263 | 0.268 | 0.181 | 0.184 |
| 1.5 | 0.598 | 0.598 | 0.589 | 0.589 | 0.531 | 0.533 | 0.392 | 0.398 | 0.258 | 0.266 | 0.177 | 0.183 |
| 2 | 0.598 | 0.598 | 0.589 | 0.589 | 0.530 | 0.532 | 0.387 | 0.397 | 0.253 | 0.265 | 0.175 | 0.182 |
| 3 | 0.598 | 0.598 | 0.589 | 0.589 | 0.528 | 0.532 | 0.381 | 0.395 | 0.249 | 0.263 | 0.172 | 0.182 |
| 4 | 0.598 | 0.598 | 0.589 | 0.589 | 0.526 | 0.531 | 0.378 | 0.394 | 0.246 | 0.263 | 0.171 | 0.181 |
| 5 | 0.598 | 0.598 | 0.589 | 0.589 | 0.526 | 0.531 | 0.376 | 0.394 | 0.245 | 0.262 | 0.171 | 0.181 |
| 7 | 0.598 | 0.598 | 0.589 | 0.589 | 0.525 | 0.531 | 0.374 | 0.393 | 0.244 | 0.262 | 0.171 | 0.182 |
| 10 | 0.598 | 0.598 | 0.589 | 0.589 | 0.524 | 0.531 | 0.374 | 0.394 | 0.244 | 0.263 | 0.171 | 0.182 |
| 15 | 0.598 | 0.598 | 0.589 | 0.589 | 0.524 | 0.531 | 0.374 | 0.394 | 0.245 | 0.264 | 0.172 | 0.182 |
| 25 | 0.598 | 0.598 | 0.589 | 0.589 | 0.524 | 0.531 | 0.377 | 0.396 | 0.248 | 0.265 | 0.173 | 0.183 |
| 50 | 0.598 | 0.598 | 0.589 | 0.589 | 0.526 | 0.532 | 0.382 | 0.398 | 0.252 | 0.267 | 0.176 | 0.184 |
| 75 | 0.598 | 0.598 | 0.589 | 0.589 | 0.527 | 0.532 | 0.385 | 0.399 | 0.255 | 0.268 | 0.178 | 0.185 |
| 100 | 0.598 | 0.598 | 0.589 | 0.589 | 0.528 | 0.532 | 0.387 | 0.400 | 0.257 | 0.269 | 0.179 | 0.185 |

The Nusselt number is plotted in the inset of Fig. 8 as a function of the Prandtl number. The figure includes data from all values of $De$ tested, for the two $\beta$ values. The heat transfer is conduction-dominated up to $Pr \approx 10^2$ ($Pe \approx 1$). In this range, the effect of $De$ and $\beta$ is relatively small and the Nusselt number tends to its theoretical value for a pure-conduction problem ($\overline{Nu} \to 2$ as $Pr \to 0$). In the convection-dominated regime, $Pr \gtrsim 10^2$, there is more dispersion of data with $De$ and $\beta$ for each $Pr$. However, it is still possible to identify a global growth tendency following approximately a power-law function. The solid line plotted in the inset of Fig. 8 represents a power-law fit to the results and its expression is given by $\overline{Nu} = 0.1109 Pr^{0.3973} + 1.765$ ($\overline{Nu} = 0.08 Pr^{0.4238} + 2$ is obtained forcing the limiting condition $\overline{Nu} \to 2$ as $Pr \to 0$). It is interesting to note that the power-law exponent is close to 1/3, which is a value reported in several correlations for spheres, concerning both Newtonian and non-Newtonian fluids (Dhole et al., 2006a, b; Finlayson and Olson, 1987; Friedlander, 1961; Westerberg and Finlayson, 1990). The difference in the scaling exponent might be attributed to the conjugate heat transfer problem solved (coupled boundary condition at the sphere surface), or simply to the absence of terms on $De$ and $\beta$ in the simple model fitted to the data.

The results for the volume-averaged temperature of the sphere are consistent with the behavior of the Nusselt number. The temperature increase is lower when the fluid is viscoelastic, displaying a minimum at the same *De* where the Nusselt number peaks.

Focusing now on a local analysis, Fig. 9 plots $Nu_\phi$ as a function of $\phi$, for $\beta = 0.1$, in the convection-dominated regime ($Pr = 10^5$). The local Nusselt number reaches its maximum in the upstream region of the sphere, and decreases as the fluid heats up and flows over the sphere surface (decreasing $\phi$). The difference between Newtonian and viscoelastic fluids also diminishes in the flow direction. The temperature at the centerline (the interval $-1 \leq x/R \leq 1$ is inside the solid sphere) is plotted in Fig. 10 together with the temperature over the sphere surface. As expected, the surface temperature increases in the flow direction, showing a rapid increase in the rear region of the sphere. At the sphere centerline, the temperature follows a quasi-parabolic-like shape, with its center shifted downstream to the sphere's center. It shall be noted that the theoretical profile expected for a pure-conduction problem ($Pr = 0$) would be a parabola centered with the sphere. In the limit $Pr \rightarrow 0$, the temperature only depends on the radial coordinate and the problem reduces to a single dimension. The detachment from this limiting case, as $Pr$ is increased, can be clearly seen in the contours of temperature plotted in Fig. 11, where there is a break of the fore-aft symmetry (characteristic of heat conduction) driven by the forced heat convection.

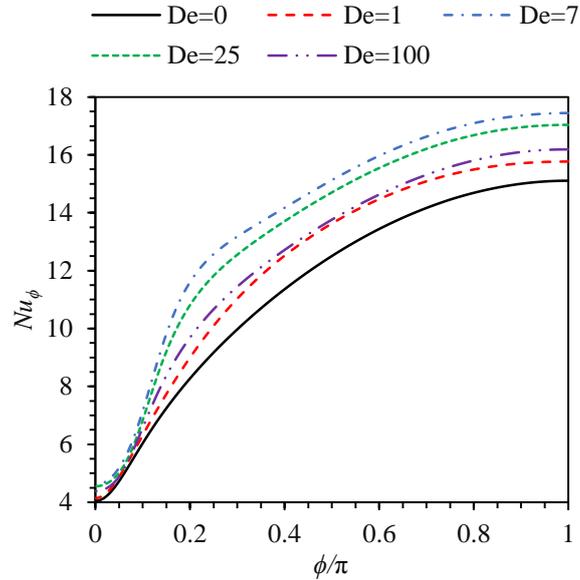

**Figure 9** – Local Nusselt number along the sphere surface for different *De* ($\beta = 0.1$, $Pr = 10^5$, $Br = 0$, $\kappa = 1$ and $\Omega = 0$). The fluid flows from $\phi/\pi = 1$ to $\phi/\pi = 0$.

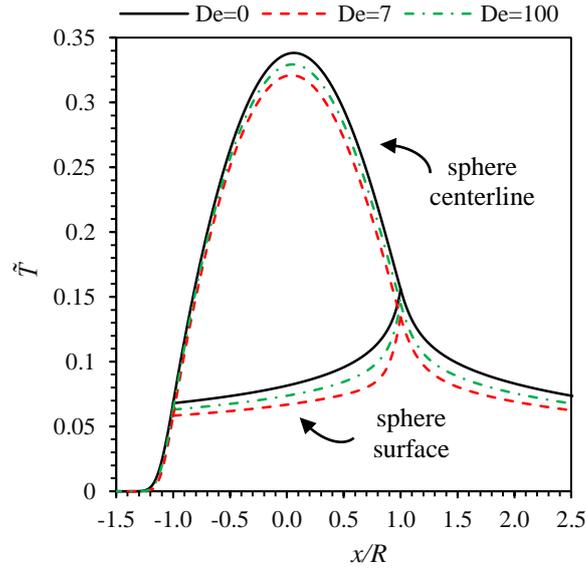

**Figure 10** – Temperature profile along the centerline and sphere surface for different $De$ ($\beta = 0.1$, $Pr = 10^5$, $Br = 0$, $\kappa = 1$ and $\Omega = 0$). For $-1 \leq x/R \leq 1$ the temperature is taken at the interior (centerline) and surface of the sphere, whereas outside of this interval the temperature is taken along the centerline in the fluid region.

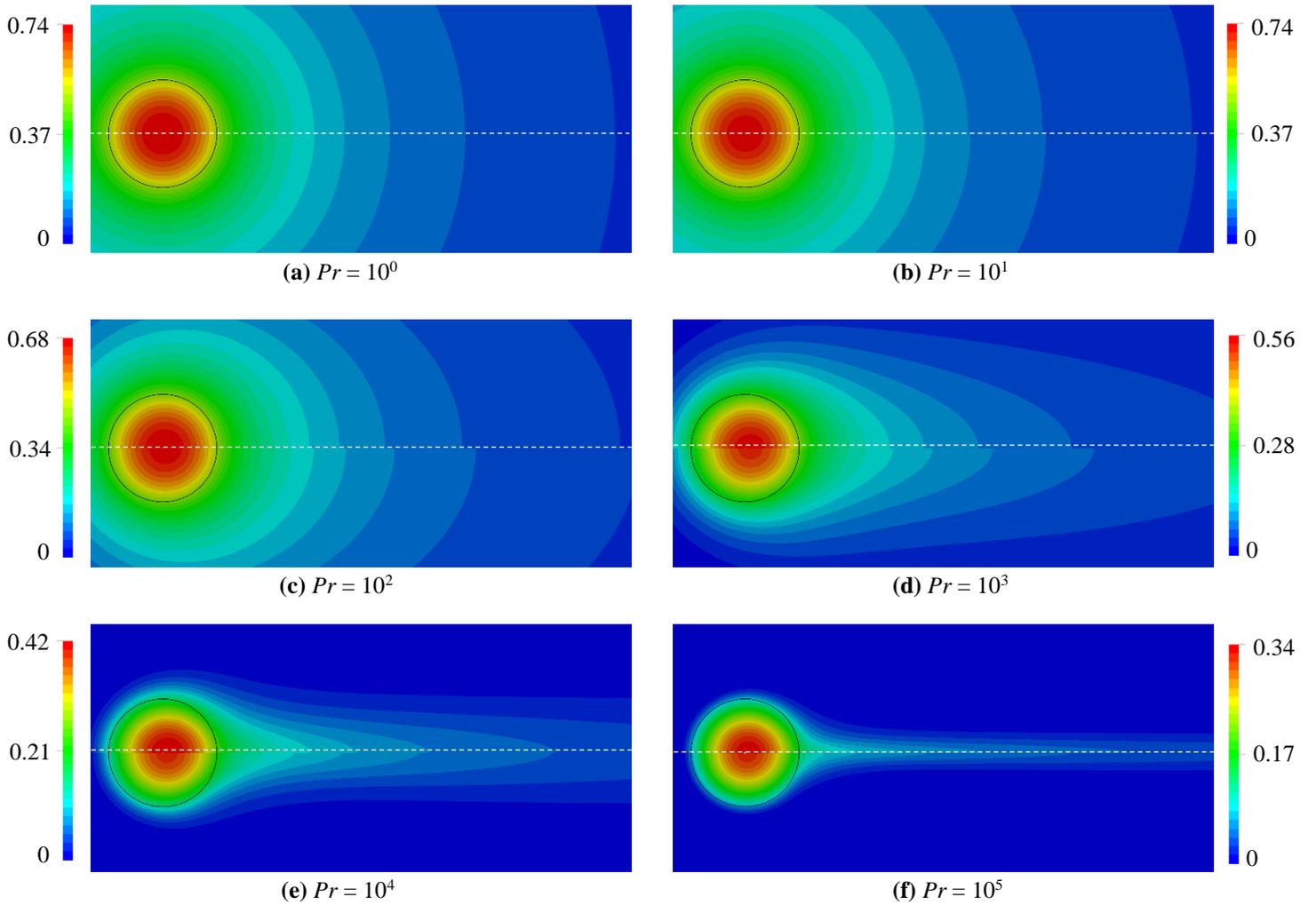

**Figure 11** – Contours of temperature ($\tilde{T}$) for different $Pr$ ($\beta = 0.1$, $Br = 0$, $\kappa = 1$ and $\Omega = 0$). In each subfigure, the upper part is for $De = 0$ (Newtonian case) and the lower part is for $De = 7$. The black solid line represents the cut of the sphere surface. The flow is from left to right.

### 6.2.2. Thermal conductivity ratio effect

The results presented in this section are for fixed $Pr = 10^5$, $Br = 0$ and $\Omega = 0$, but varying $\kappa$. The high Prandtl (and Péclet) number was selected (and kept in the remaining sections) to ensure a convection-dominated heat transfer regime, where the effect of $De$ is more significant, as seen in the previous section. The simulations with $\kappa = 1$ were conducted in the previous section and we now consider $\kappa = 0.1$ and $\kappa = 10$, in order to assess the effect of this parameter on heat transfer.

The surface-averaged Nusselt number and volume-averaged sphere temperature are presented in Table 5. The differences in the average Nusselt number are small among the three values of $\kappa$ tested, although it is clear that a lower $\kappa$ improves the heat transfer, probably due to an increase of the driving temperature difference ($T_s$–$T_f$) induced from the sphere side. The effect of $\kappa$ is more evident in the volume-averaged temperature of the sphere, with a clear increase of the dimensionless temperature as the thermal conduction inside the sphere becomes less significant ($\kappa$ decreases). Again, the fluid with lower $\beta$ displays a higher heat transfer coefficient.

The Biot number dependence on $\kappa$ seems to closely follow a power-law function, $Bi = \alpha \kappa^m$, with the exponent $m$ close to -1 and nearly independent from $De$ and $\beta$. A set of simulations was conducted for $Pr = 10^0, 10^1, 10^2, 10^3$ and $10^4$ (results not shown in Table 5) and it was observed that the power-law evolution is still valid, with $m \rightarrow$ -1 as $Pr \rightarrow 0$, and $\alpha$ being essentially a function of $Pr$. This is consistent with the invariance of $\overline{Nu}$ as $Pr \rightarrow 0$, as we recall that $\overline{Nu} = 6\kappa Bi = 6\alpha\kappa^{m+1}$.

The temperature profiles are continuous at the centerline, but the temperature gradient is different on each side of the interface, as required to ensure the continuity of the heat flux when there is a mismatch in the thermal conductivities (Fig. 12). Moreover, a higher value of $\kappa$ (lower $Bi$) leads to a more homogeneous distribution of temperature inside the sphere and increases the deviation of the temperature profile at the centerline from the parabolic shape (Fig. 12).

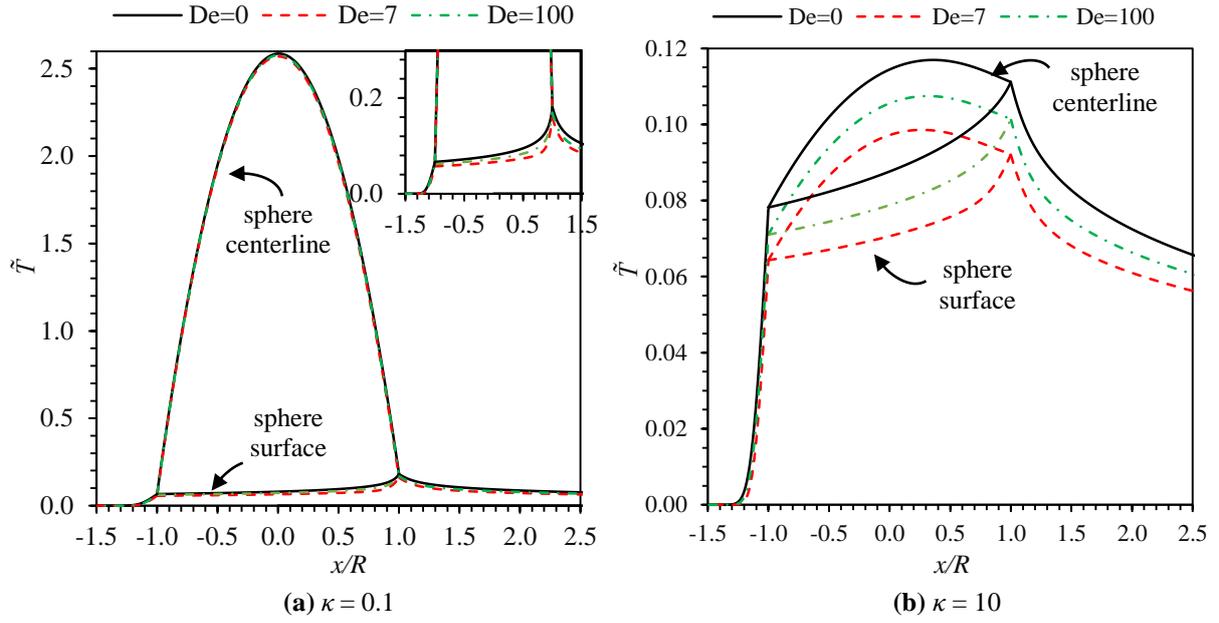

**Figure 12** – Temperature profile along the centerline and sphere surface for (a) $\kappa = 0.1$ and (b) $\kappa = 10$ ($\beta = 0.1$, $Pr = 10^5$, $Br = 0$ and $\Omega = 0$). For $-1 \leq x/R \leq 1$ the temperature is taken at the interior (centerline) and surface of the sphere, whereas outside of this interval the temperature is taken along the centerline in the fluid region. The inset in (a) is a zoomed view of the main figure.

**Table 5** – Surface-averaged Nusselt number and volume-averaged sphere temperature for different $De$, $\beta$ and $\kappa$ ($Pr = 10^5$, $Br = 0$ and $\Omega = 0$). Note that the data for $\kappa = 1$ is repeated here from Tables 3 and 4 for the ease of comparison.

| | $\beta = 0.1$ | | | | | | $\beta = 0.5$ | | | | | |
|---|---|---|---|---|---|---|---|---|---|---|---|---|
| | $\kappa = 0.1$ | | $\kappa = 1$ | | $\kappa = 10$ | | $\kappa = 0.1$ | | $\kappa = 1$ | | $\kappa = 10$ | |
| $De$ | $\overline{Nu}$ | $\overline{\tilde{T}}_s$ | $\overline{Nu}$ | $\overline{\tilde{T}}_s$ | $\overline{Nu}$ | $\overline{\tilde{T}}_s$ | $\overline{Nu}$ | $\overline{\tilde{T}}_s$ | $\overline{Nu}$ | $\overline{\tilde{T}}_s$ | $\overline{Nu}$ | $\overline{\tilde{T}}_s$ |
| 0 | 12.044 | 1.087 | 11.956 | 0.187 | 11.454 | 0.099 | 12.044 | 1.087 | 11.956 | 0.187 | 11.454 | 0.099 |
| 0.1 | 12.059 | 1.087 | 11.971 | 0.187 | 11.467 | 0.099 | 12.053 | 1.087 | 11.965 | 0.187 | 11.462 | 0.099 |
| 0.5 | 12.370 | 1.085 | 12.284 | 0.185 | 11.770 | 0.097 | 12.205 | 1.086 | 12.118 | 0.186 | 11.611 | 0.098 |
| 1 | 12.981 | 1.080 | 12.900 | 0.181 | 12.387 | 0.093 | 12.433 | 1.084 | 12.348 | 0.184 | 11.843 | 0.096 |
| 1.5 | 13.512 | 1.077 | 13.437 | 0.177 | 12.938 | 0.089 | 12.592 | 1.083 | 12.508 | 0.183 | 12.007 | 0.095 |
| 2 | 13.889 | 1.074 | 13.818 | 0.175 | 13.333 | 0.087 | 12.691 | 1.082 | 12.608 | 0.182 | 12.110 | 0.094 |
| 3 | 14.318 | 1.072 | 14.249 | 0.172 | 13.775 | 0.084 | 12.788 | 1.081 | 12.706 | 0.182 | 12.211 | 0.094 |
| 4 | 14.519 | 1.071 | 14.451 | 0.171 | 13.977 | 0.083 | 12.823 | 1.081 | 12.740 | 0.181 | 12.246 | 0.093 |
| 5 | 14.616 | 1.070 | 14.548 | 0.171 | 14.071 | 0.082 | 12.830 | 1.081 | 12.748 | 0.181 | 12.254 | 0.093 |
| 7 | 14.672 | 1.070 | 14.603 | 0.171 | 14.122 | 0.082 | 12.814 | 1.081 | 12.732 | 0.182 | 12.237 | 0.094 |
| 10 | 14.628 | 1.071 | 14.559 | 0.171 | 14.073 | 0.082 | 12.768 | 1.081 | 12.685 | 0.182 | 12.190 | 0.094 |
| 15 | 14.479 | 1.071 | 14.409 | 0.172 | 13.921 | 0.083 | 12.691 | 1.082 | 12.607 | 0.182 | 12.111 | 0.094 |
| 25 | 14.195 | 1.073 | 14.123 | 0.173 | 13.634 | 0.085 | 12.575 | 1.083 | 12.490 | 0.183 | 11.992 | 0.095 |
| 50 | 13.730 | 1.075 | 13.654 | 0.176 | 13.162 | 0.088 | 12.416 | 1.084 | 12.330 | 0.184 | 11.831 | 0.097 |
| 75 | 13.448 | 1.077 | 13.370 | 0.178 | 12.877 | 0.089 | 12.334 | 1.085 | 12.248 | 0.185 | 11.748 | 0.097 |
| 100 | 13.267 | 1.078 | 13.188 | 0.179 | 12.693 | 0.091 | 12.283 | 1.085 | 12.197 | 0.185 | 11.696 | 0.098 |

### 6.2.3. Thermal contact resistance effect

The results presented in this section are for fixed $Pr = 10^5$, $Br = 0$ and $\kappa = 1$, but varying $\Omega$. Three values of $\Omega$ were tested, $\Omega = 0.05$, 0.1 and 0.5. We remember that the case $\Omega = 0$ represents perfect thermal contact and has been addressed in the previous sections.

The surface-averaged Nusselt number and volume-averaged sphere temperature are listed in Table 6 for the different combinations of $De$ and $\beta$ tested. The higher the contact resistance, the higher the sphere temperature, as expected. Nonetheless, there is only a very slight variation of the average Nusselt number – $\overline{Nu}$ only increases marginally with $\Omega$ (Table 6).

**Table 6** – Surface-averaged Nusselt number and volume-averaged sphere temperature for different $De$, $\beta$ and $\Omega$ ($Pr = 10^5$, $Br = 0$ and $\kappa = 1$).

| | $\beta = 0.1$ | | | | | | $\beta = 0.5$ | | | | | |
|---|---|---|---|---|---|---|---|---|---|---|---|---|
| $De$ | $\Omega = 0.05$ | | $\Omega = 0.1$ | | $\Omega = 0.5$ | | $\Omega = 0.05$ | | $\Omega = 0.1$ | | $\Omega = 0.5$ | |
| | $\overline{Nu}$ | $\overline{\widetilde{T}}_s$ | $\overline{Nu}$ | $\overline{\widetilde{T}}_s$ | $\overline{Nu}$ | $\overline{\widetilde{T}}_s$ | $\overline{Nu}$ | $\overline{\widetilde{T}}_s$ | $\overline{Nu}$ | $\overline{\widetilde{T}}_s$ | $\overline{Nu}$ | $\overline{\widetilde{T}}_s$ |
| 0 | 11.962 | 0.237 | 11.968 | 0.287 | 11.998 | 0.687 | 11.962 | 0.237 | 11.968 | 0.287 | 11.998 | 0.687 |
| 0.1 | 11.978 | 0.237 | 11.983 | 0.287 | 12.013 | 0.687 | 11.972 | 0.237 | 11.977 | 0.287 | 12.007 | 0.687 |
| 0.5 | 12.290 | 0.235 | 12.295 | 0.285 | 12.324 | 0.685 | 12.124 | 0.236 | 12.130 | 0.286 | 12.159 | 0.686 |
| 1 | 12.906 | 0.231 | 12.911 | 0.281 | 12.937 | 0.681 | 12.354 | 0.234 | 12.360 | 0.284 | 12.388 | 0.684 |
| 1.5 | 13.441 | 0.227 | 13.446 | 0.277 | 13.470 | 0.677 | 12.514 | 0.233 | 12.519 | 0.283 | 12.547 | 0.683 |
| 2 | 13.822 | 0.225 | 13.826 | 0.275 | 13.849 | 0.675 | 12.614 | 0.232 | 12.619 | 0.282 | 12.646 | 0.682 |
| 3 | 14.253 | 0.222 | 14.257 | 0.272 | 14.279 | 0.672 | 12.712 | 0.232 | 12.717 | 0.282 | 12.744 | 0.682 |
| 4 | 14.455 | 0.221 | 14.458 | 0.271 | 14.480 | 0.671 | 12.746 | 0.231 | 12.752 | 0.281 | 12.779 | 0.681 |
| 5 | 14.551 | 0.221 | 14.555 | 0.271 | 14.577 | 0.671 | 12.754 | 0.231 | 12.759 | 0.281 | 12.786 | 0.681 |
| 7 | 14.607 | 0.221 | 14.610 | 0.271 | 14.632 | 0.671 | 12.738 | 0.232 | 12.743 | 0.282 | 12.770 | 0.681 |
| 10 | 14.562 | 0.221 | 14.566 | 0.271 | 14.588 | 0.671 | 12.691 | 0.232 | 12.697 | 0.282 | 12.724 | 0.682 |
| 15 | 14.413 | 0.222 | 14.417 | 0.272 | 14.439 | 0.672 | 12.613 | 0.232 | 12.619 | 0.282 | 12.647 | 0.682 |
| 25 | 14.127 | 0.223 | 14.131 | 0.273 | 14.155 | 0.673 | 12.496 | 0.233 | 12.502 | 0.283 | 12.530 | 0.683 |
| 50 | 13.659 | 0.226 | 13.663 | 0.276 | 13.688 | 0.676 | 12.336 | 0.234 | 12.342 | 0.284 | 12.371 | 0.684 |
| 75 | 13.375 | 0.228 | 13.380 | 0.278 | 13.405 | 0.677 | 12.254 | 0.235 | 12.260 | 0.285 | 12.288 | 0.685 |
| 100 | 13.193 | 0.229 | 13.198 | 0.279 | 13.224 | 0.679 | 12.203 | 0.235 | 12.209 | 0.285 | 12.238 | 0.685 |

The temperature profiles along the centerline are discontinuous at the solid-fluid interface, presenting a different value on each side, but equal derivative (Fig. 13). The temperature profiles in the fluid side (centerline and sphere surface) change very slightly with $\Omega$. Indeed, at the plotting scale of Fig. 13 it is difficult to distinguish these profiles for the different values of $\Omega$. On the other hand, the effect of $\Omega$ on the temperature profiles in the solid side (centerline and sphere surface) is more notorious (Fig. 13). These profiles are vertically shifted as $\Omega$ increases, with only minor changes of shape at the rear of the sphere.

Similar to the effect observed for $\kappa$, varying the value of $\Omega$ does not influence the non-monotonic behavior of the $\overline{Nu} - De$ relation, and the maximum is still reached at the same $De$. The more elastic fluid ($\beta = 0.1$) again shows a stronger heat transfer enhancement.

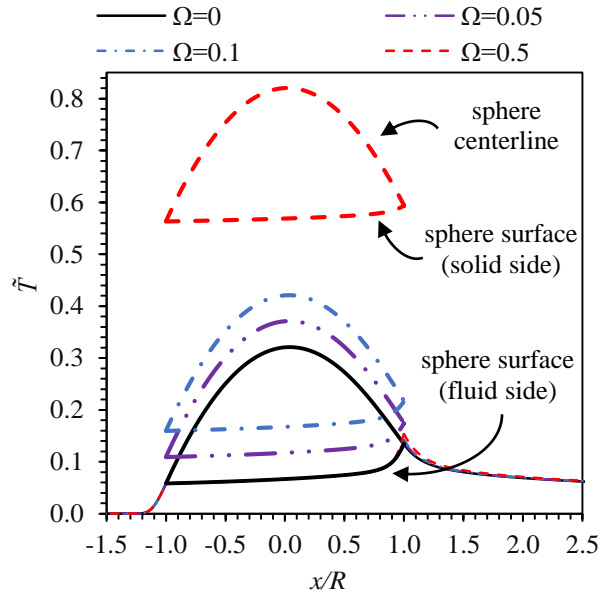

**Figure 13** – Temperature profile along the centerline and sphere surface for four different values of the contact resistance ($De = 7$, $\beta = 0.1$, $Pr = 10^5$, $Br = 0$ and $\kappa = 1$). In the range $-1 \leq x/R \leq 1$, three different profiles are plotted: the temperature along the sphere centerline (thick lines), the temperature along the sphere surface on the solid side (thick lines) and the temperature along the sphere surface on the fluid side (thin lines). For $\Omega = 0$, the profiles at the sphere surface on the solid and fluid side coincide with each other. Note that the profiles of temperature over the sphere surface on the fluid side are visually indistinguishable from each other.

### 6.2.4. Viscous dissipation effect

The results presented in this section are for fixed $Pr = 10^5$, $\kappa = 1$ and $\Omega = 0$, but varying $Br$. This last set of simulations was performed taking into account viscous dissipation, i.e. $Br \neq 0$. Under these conditions, in addition to the heat generated inside the sphere, there is also heat generation in the fluid due to viscous dissipation. The simulations were run for $Br = 1$, 10 and 100.

The heat transfer enhancement relative to the Newtonian case is plotted in Figs. 14(a) and (b) for the several conditions tested. While the $\overline{Nu} - De$ relation is non-monotonic at low $Br$, there is an inversion of behavior at $Br \approx 1$; the curves for $Br = 10$ and $Br = 100$ show a continuous increase of $\overline{Nu}$ with $De$. This is accompanied by a continuous decrease of the average temperature of the sphere with *De* at such high *Br* (Table 7). As *Br* increases, the local Nusselt number decreases over the whole surface of the sphere, even becoming negative in the rear region of the sphere ($\phi \lesssim 57°$) for *Br* = 100 (Fig. 14c). The local temperature profile also suffers changes with *Br*, as shown in Fig. 14(d). While the maximum temperature along the centerline occurs inside the sphere at low *Br*, the peak temperature shifts to the fluid region in the rear of the sphere as *Br* is increased to 100.

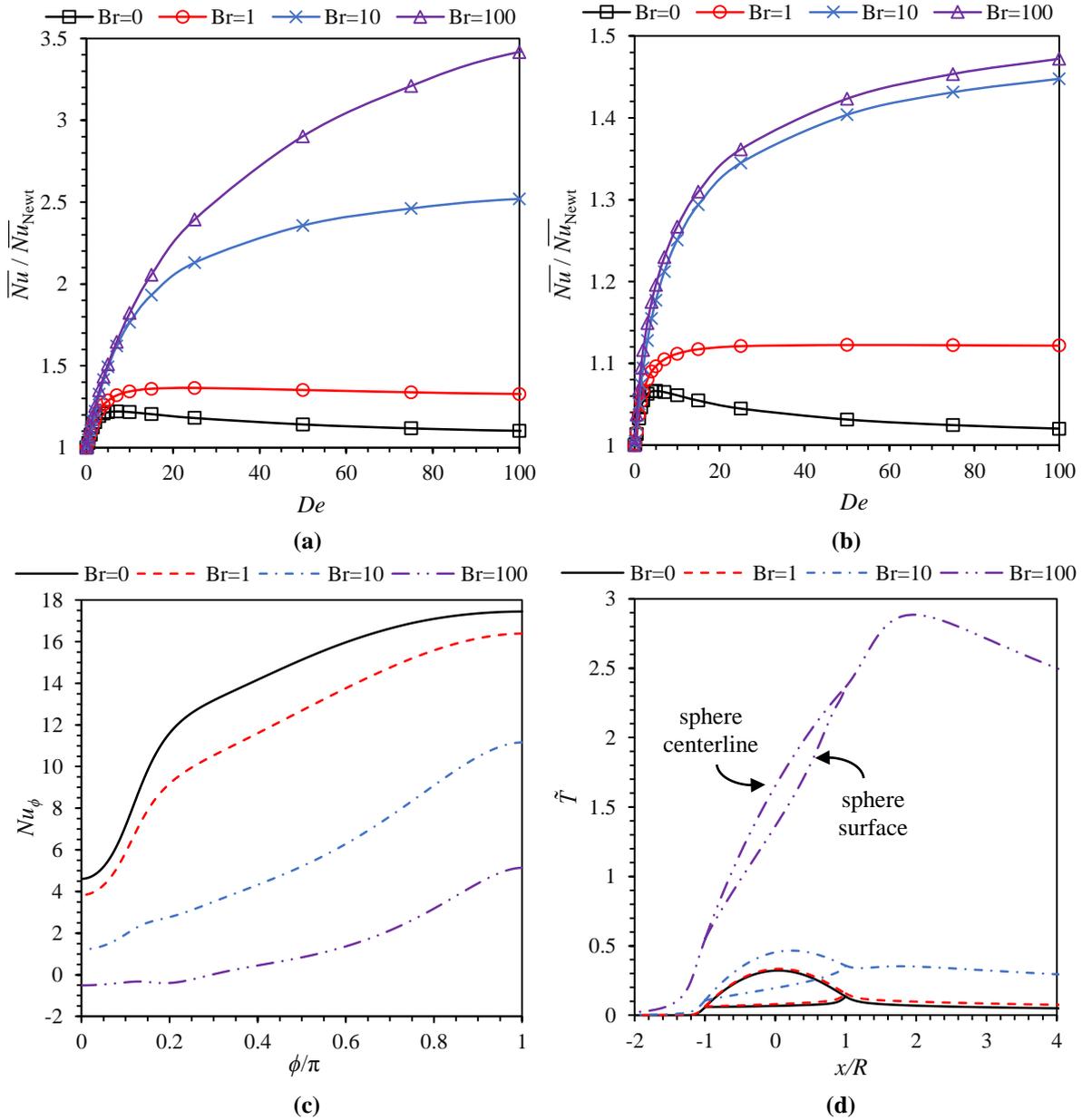

**Figure 14** – Viscous dissipation effect on heat transfer: (a) Nusselt number ratio for $\beta = 0.1$; (b) Nusselt number ratio for $\beta = 0.5$; (c) local Nusselt number profile along the sphere surface for $\beta = 0.1$ and *De* = 7; (d) temperature profile along the centerline and sphere surface for $\beta = 0.1$ and *De* = 7 (in the range $-1 \leq x/R \leq 1$ the temperature is taken along the centerline and surface of the sphere, whereas outside this interval the temperature is taken along the centerline in the fluid region). In (a) and (b), the lines are only a guide to the eye. All the results plotted were obtained for $Pr = 10^5$, $\kappa = 1$ and $\Omega = 0$.

These results can be explained by distinguishing between the range of low *Br* and the range of high *Br*. For low *Br,* the heat generated by viscous dissipation is negligible compared to the heat generated by the sphere. The sphere transfers heat to the fluid over its whole surface. The layer of fluid surrounding the sphere is colder than the surface of the sphere, but as *Br* increases this layer becomes warmer due to viscous dissipation, and the average Nusselt number decreases. For high *Br*, viscous dissipation generates larger amounts of heat and there are regions where the fluid layer adjacent to sphere is hotter than the sphere surface. In those regions, the sphere receives heat from the fluid and $Nu_\phi$ becomes locally negative. In the limit, when $Br \rightarrow +\infty$ the heat generated by the sphere is negligible and all the heat transferred arises from viscous dissipation. If further $Pr \rightarrow 0$, then $\overline{Nu} \rightarrow 0$, which contrasts with the limit $\overline{Nu} \rightarrow 2$ at low *Br*.

**Table 7** – Surface-averaged Nusselt number and volume-averaged sphere temperature for different *De, β* and *Br* ($Pr = 10^5$, $\kappa = 1$ and $\Omega = 0$).

| De | β = 0.1 ||||||  β = 0.5 ||||||
| | Br = 1 || Br = 10 || Br = 100 || Br = 1 || Br = 10 || Br = 100 ||
| | $\overline{Nu}$ | $\overline{\widetilde{T}_s}$ | $\overline{Nu}$ | $\overline{\widetilde{T}_s}$ | $\overline{Nu}$ | $\overline{\widetilde{T}_s}$ | $\overline{Nu}$ | $\overline{\widetilde{T}_s}$ | $\overline{Nu}$ | $\overline{\widetilde{T}_s}$ | $\overline{Nu}$ | $\overline{\widetilde{T}_s}$ |
|---|---|---|---|---|---|---|---|---|---|---|---|---|
| 0 | 9.429 | 0.212 | 3.439 | 0.434 | 0.700 | 2.649 | 9.429 | 0.212 | 3.439 | 0.434 | 0.700 | 2.649 |
| 0.1 | 9.459 | 0.212 | 3.470 | 0.431 | 0.711 | 2.629 | 9.445 | 0.212 | 3.455 | 0.432 | 0.706 | 2.639 |
| 0.5 | 9.730 | 0.209 | 3.600 | 0.422 | 0.748 | 2.553 | 9.580 | 0.210 | 3.523 | 0.427 | 0.726 | 2.599 |
| 1 | 10.224 | 0.203 | 3.798 | 0.407 | 0.792 | 2.441 | 9.777 | 0.208 | 3.612 | 0.420 | 0.748 | 2.543 |
| 1.5 | 10.682 | 0.199 | 4.006 | 0.392 | 0.836 | 2.321 | 9.929 | 0.206 | 3.694 | 0.414 | 0.766 | 2.488 |
| 2 | 11.048 | 0.195 | 4.204 | 0.378 | 0.876 | 2.204 | 10.040 | 0.205 | 3.764 | 0.408 | 0.781 | 2.437 |
| 3 | 11.564 | 0.191 | 4.561 | 0.355 | 0.944 | 1.996 | 10.185 | 0.203 | 3.879 | 0.398 | 0.804 | 2.347 |
| 4 | 11.902 | 0.188 | 4.868 | 0.337 | 1.002 | 1.827 | 10.276 | 0.202 | 3.971 | 0.391 | 0.822 | 2.273 |
| 5 | 12.138 | 0.186 | 5.135 | 0.323 | 1.056 | 1.695 | 10.339 | 0.202 | 4.048 | 0.384 | 0.837 | 2.212 |
| 7 | 12.436 | 0.184 | 5.574 | 0.304 | 1.152 | 1.501 | 10.418 | 0.201 | 4.169 | 0.375 | 0.861 | 2.117 |
| 10 | 12.666 | 0.182 | 6.070 | 0.285 | 1.275 | 1.313 | 10.482 | 0.200 | 4.300 | 0.365 | 0.887 | 2.016 |
| 15 | 12.819 | 0.181 | 6.642 | 0.267 | 1.439 | 1.128 | 10.534 | 0.200 | 4.449 | 0.355 | 0.917 | 1.908 |
| 25 | 12.868 | 0.181 | 7.326 | 0.250 | 1.675 | 0.938 | 10.570 | 0.199 | 4.625 | 0.344 | 0.953 | 1.786 |
| 50 | 12.747 | 0.182 | 8.106 | 0.233 | 2.031 | 0.751 | 10.583 | 0.199 | 4.827 | 0.332 | 0.996 | 1.655 |
| 75 | 12.614 | 0.183 | 8.464 | 0.227 | 2.246 | 0.671 | 10.580 | 0.199 | 4.922 | 0.326 | 1.017 | 1.596 |
| 100 | 12.515 | 0.183 | 8.666 | 0.224 | 2.391 | 0.627 | 10.576 | 0.199 | 4.979 | 0.323 | 1.030 | 1.562 |

The contours plotted in Fig. 15 offer a global view of the temperature distribution inside the sphere and in the fluid for *De* = 7. With the increase of *Br*, the sphere becomes hotter on the downstream side due to the cumulative effect of viscous dissipation. The isotherms inside the sphere are no longer spherical, but become nearly normal to the axis of symmetry (flow direction). In the fluid region, the volume of fluid verifying $\tilde{T} > \delta$, with $\delta > 0$, increases with $Br$ due to viscous dissipation.

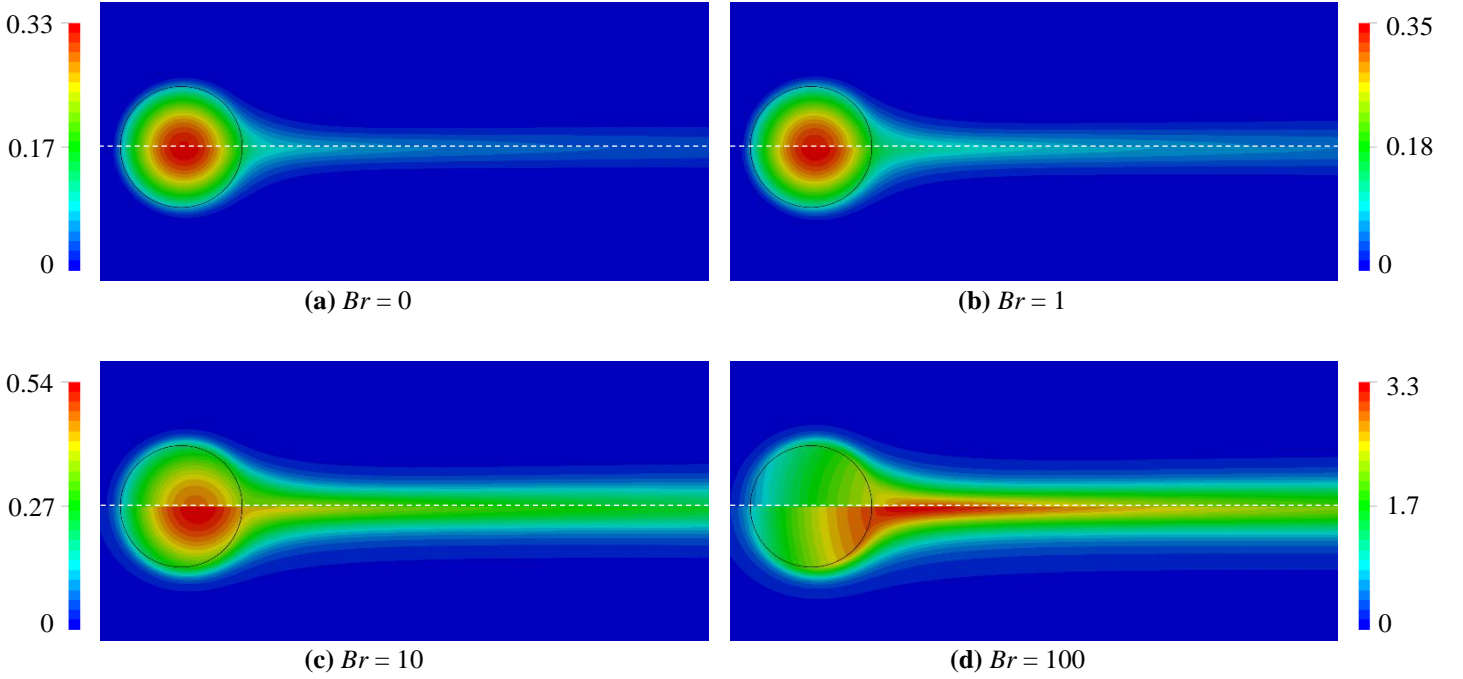

**Figure 15** – Contours of temperature ($\tilde{T}$) for different $Br$ values ($De = 7$, $Pr = 10^5$, $\kappa = 1$ and $\Omega = 0$). In each subfigure, the upper part is for $\beta = 0.1$ and the lower part is for $\beta = 0.5$. The black solid line represents the cut of the sphere surface. The flow is from left to right.

### 6.2.5. Relation between the average sphere temperature and Biot number

In this section, we aim to find a correlation between the volume-averaged temperature of the sphere and the Biot number for all data obtained in the absence of viscous dissipation ($Br = 0$). Since the Biot number is related with the solid body, the dimensionless volume-averaged temperature of the sphere, defined in Eq. (16), was rescaled to incorporate the thermal conductivity of the sphere, i.e. we defined a new temperature variable $\overline{\tilde{T}_s^*} = \kappa \overline{\tilde{T}_s}$. The plot of $\overline{\tilde{T}_s^*}$ as a function of $Bi$ is presented in Fig. 16, which includes all the data obtained in the previous sections for $Br = 0$, listed in Tables 3–6 (open symbols). Some additional simulations were carried out for $\kappa = 0.1$ and $\kappa = 10$ at $Pr = 10^0, 10^1, 10^2, 10^3$ and $10^4$ to further populate the master curve of Fig. 16. These data points are represented as filled symbols in the curve and they fall in the ranges $0.03 < Bi < 0.12$ and $3 < Bi < 12$. The relation between $\overline{\tilde{T}_s^*}$ and $Bi$ is well evidenced by the master curve, which can be approximated as $\overline{\tilde{T}_s^*} = 0.1672 Bi^{-0.9989} + 0.1023$ with a maximum error below 2 %. This expression was obtained from data in the range: $0 \leq De \leq 100$, $10^0 \leq Pr \leq 10^5$, $0.1 \leq \kappa \leq 10$, $0 \leq \Omega \leq 0.5$ and $\beta = \{0.1, 0.5\}$, for fixed $Re = 0.01$, $\varepsilon$

= 0.25 and $Br = 0$. However, we expect this expression to keep valid for wider ranges of the variable parameters and, eventually, for different values of the fixed parameters. The data obtained for $Br \neq 0$ still aligns with the master curve for low values of $Br$, but starts to deviate as $Br$ increases ($Br \gtrsim 1$), calling for the need of a more complex expression incorporating the effect of $Br$.

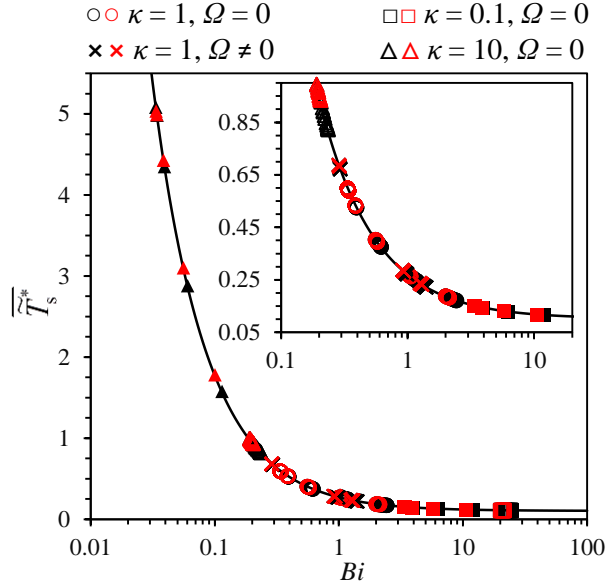

**Figure 16** – Master curve for the rescaled temperature ($\overline{\widetilde{T}_s^*}$) as a function of Biot number. All data points obtained in this work in the absence of viscous dissipation are plotted here. Black symbols represent data for $\beta = 0.1$ and red symbols are for $\beta = 0.5$. Open symbols are for data listed in Tables 3–6, whereas filled symbols represent non-listed data (see the text). The solid line represents a power-law fit to the data and the inset is a zoomed view of the main figure.

## 7. Conclusions

The heat transfer and unbounded flow of a simplified PTT fluid past a sphere was investigated in this work. The sphere generates heat in its interior at a constant and uniform rate. Keeping the Reynolds number fixed ($Re = 0.01$), we investigated the effect of Prandtl number, Brinkman number, thermal conductivity ratio and thermal contact resistance for Deborah numbers in the interval $0 \leq De \leq 100$ and two different solvent viscosity ratios ($\beta = 0.1$ and 0.5). The assumption of temperature-independent parameters allowed decoupling the flow from the heat transfer process.

The drag coefficient of the sphere showed a monotonic decrease with $De$, attributed to the shear-thinning behavior of the PTT model. The fluid with lower $\beta$ presented a lower drag coefficient. The stresses acting on the sphere surface and in the wake of the sphere decrease with $De$ after an initial increase up to a critical $De$. A negative wake was observed downstream of the sphere for both values of $\beta$ after exceeding a threshold $De$, being more intense for the more elastic fluid (lower $\beta$).

In the absence of viscous dissipation, the Nusselt number increases with *De* up to a critical *De*, above which it decreases. The more elastic fluid (*β* = 0.1) displays a higher Nusselt number and the average dimensionless temperature of the sphere is consequently lower. The Nusselt number scales proportionally to $Pr^m$, with $m \approx 0.4$, and a power-law relation was also found between the Biot number and the volume-averaged temperature of the sphere, which kept valid for the simulations with variable conductivity ratio and contact resistance.

A higher ratio of thermal conductivities (solid more conductive than fluid) lowers and homogenizes the temperature of the sphere, and decreases simultaneously the Nusselt and Biot numbers.

The presence of a thermal contact resistance at the solid-fluid interface increases the temperature on the solid side. As the resistance increases, the temperature profiles on the solid side are shifted to higher values, but their shape only suffers minor changes. The effect on the Nusselt number is small.

The introduction of viscous dissipation in the fluid can drastically change the heat transfer process. At sufficiently high Brinkman numbers, part of the sphere receives heat from the fluid.

The present work can be complemented in several ways by future studies. One possible direction is to analyze the effect of the parameters that were held fixed in this work: the Reynolds number and the extensibility parameter of the PTT model, possibly accounting for temperature-dependent properties. Other variables can be added to the problem, as for example slip on the sphere surface and natural convection. Another possible direction is the investigation of the transients of this problem. Due to the scarcity of results on the heat transfer problem around a sphere involving viscoelastic fluids, studying the fluid side in isolation, by imposing a fixed heat flux or fixed temperature on the sphere surface, is still another possible direction of research.

**Appendix A**

Discrete numerical methods like finite-volumes present, in general, a dependency of the solution on the spatial resolution of the computational grid. Besides, the problem under study has additional dependency on the size of the simulation domain due to the artificial boundary conditions applied at the surrounding boundaries to simulate an unbounded flow. This last question was discussed several times in the literature and domains of different sizes have been employed in different works (Faroughi et al., 2020;

Kishore and Ramteke, 2016). The final error affecting the solution depends on many factors, such as the Reynolds number, the nature of the fluid (Newtonian *vs.* non-Newtonian), the type of artificial boundary conditions applied on the outer boundary (outflow, symmetry, periodicity, etc.), the type of numerical methods employed, among others. The error also affects differently each post-processing variable. Therefore, the existence of an optimal domain size and mesh resolution for all conditions is unrealistic if computational cost is to be taken into account. Instead, these parameters should be selected individually for each specific problem as a compromise between accuracy and computational cost.

In order to assess the mesh and domain size dependency of the solution, a set of simulations was carried out in grids of different resolutions and sizes. Keeping the domain size fixed at $200R$, the mesh resolution was varied as indicated in Table 1, originating meshes M1, M2 and M3. A second set of meshes with fixed resolution but different size was generated by varying the domain size of mesh M2, originating meshes $M2_{R50}$, $M2_{R100}$, $M2_{R150}$, $M2_{R250}$, $M2_{R300}$ and $M2_{R350}$ (it is implicit that M2 is in fact $M2_{R200}$). As the names suggest, the size of the domain in these meshes varies between $50R$ and $350R$ (the number of cells in the fluid side increases with domain size, since the resolution is kept constant). The simulations were carried out for $De = 10$, $\beta = 0.5$, $Pr = 10^5$, $Br = 0$, $\kappa = 1$ and $\Omega = 0$.

The results obtained are listed in Table I. It can be shown that the mesh resolution has a small effect on the parameters under analysis, which is indicative that the range of mesh resolutions tested is already appropriate to capture accurately the main characteristics of the flow, namely the thin thermal boundary layer formed around the sphere. On the other hand, the domain size seems to have a more important influence on the solution, for this low *Re* flow, vanishing as the domain size increases. However, a larger domain leads to a higher computational cost, not only because of the higher number of cells in the mesh, but also due to the higher number of iterations needed to converge to steady-state. Based on these results, mesh M2 was selected and used throughout this work as a compromise between accuracy and computational cost (each simulation of the flow in mesh M2 takes approximately 24 h to complete using 7 processors; the heat transfer simulation only takes a few minutes for each set of conditions).

It is worth mentioning that when the Oldroyd-B or Upper-Convected Maxwell models are used, the stress profiles typically show a high dependency on the mesh resolution in the wake of the sphere for increasing *De* (Faroughi et al., 2020). This issue is common to the simulation of flow around cylinders with Upper-Convected Maxwell and Oldroyd-B fluids (Afonso et al., 2008; Alves et al., 2001), where both type of fluid flows form an intense and thin birefringent strand in the wake of the obstacle, due to the growth of normal stresses with *De*. This is not always reflected in the drag coefficient, or at least not with the same intensity, as the stresses acting on the sphere are typically less dependent on mesh resolution. In our case, this issue does not arise because the stresses decrease in the wake of the sphere with the increase of *De* (Figs. 6c and d). We verified the mesh independency of the stresses profiles at different *De* values and it was confirmed that such profiles are in fact mesh-independent for all the resolutions tested (M1, M2 and M3).

**Table I** – Effect of mesh resolution and size of the simulation domain on the drag coefficient, surface-averaged Nusselt number, volume-averaged and maximum sphere temperature. The simulations were performed for $De = 10$, $\beta = 0.5$, $Pr = 10^5$, $Br = 0$, $\kappa = 1$ and $\Omega = 0$.

| Mesh | $C_d Re$ | $\overline{Nu}$ | $\overline{\tilde{T}_S}$ | $\tilde{T}_{S,max}$ |
|---|---|---|---|---|
| M1 | 17.1438 | 12.6844 | 0.1819 | 0.3325 |
| M2 | 17.1458 | 12.6852 | 0.1819 | 0.3325 |
| M3 | 17.1500 | 12.6856 | 0.1819 | 0.3325 |
| M2$_{R50}$ | 17.4603 | 12.7673 | 0.1814 | 0.3320 |
| M2$_{R100}$ | 17.2452 | 12.7137 | 0.1817 | 0.3324 |
| M2$_{R150}$ | 17.1752 | 12.6944 | 0.1818 | 0.3325 |
| M2$_{R250}$ | 17.1268 | 12.6799 | 0.1819 | 0.3326 |
| M2$_{R300}$ | 17.1143 | 12.6764 | 0.1820 | 0.3326 |
| M2$_{R350}$ | 17.1056 | 12.6740 | 0.1820 | 0.3326 |